\documentclass[12pt]{iopart}

\usepackage{color}
\usepackage{ulem}
\usepackage{cancel}
\usepackage{graphicx}

\begin{document}

\title{Thermo-elastic induced phase noise in the LISA Pathfinder spacecraft}

\author{F Gibert$^1$, M Nofrarias$^1$, N Karnesis$^1$, L Gesa$^1$, V Mart\' in$^1$, \\ I Mateos$^1$, A Lobo$^1$\footnote{Deceased}, R Flatscher$^2$, D Gerardi$^2$, J Burkhardt$^2$, R Gerndt$^2$, D I Robertson$^3$, H Ward$^3$, P W McNamara$^4$, F Guzm\' an$^5$, M~Hewitson$^5$, I Diepholz$^5$, J Reiche$^5$, G Heinzel$^5$, K~Danzmann$^5$}

\address{$^1$ Institut de Ci\`encies de l'Espai (ICE-CSIC/IEEC), Barcelona, Spain}

\address{$^2$ Airbus Defence and Space GmbH, Friedrichshafen, Germany}

\address{$^3$ Institute for Gravitational Research, SUPA, Glasgow, UK}

\address{$^4$ European Space Research and Technology Centre, Noordwijk, The Netherlands}

\address{$^5$ Max-Planck-Institut fur Gravitationsphysik, Hannover, Germany}

\ead{gibert@ieec.cat}
\begin{abstract}
During the On-Station Thermal Test campaign of the LISA Pathfinder the Diagnostics Subsystem was tested in nearly space conditions for the first time after integration in the satellite. The results showed the compliance of the temperature measurement system, obtaining temperature noise around $10^{-4}\,{\rm K}\, {\rm Hz}^{-1/2}$ in the frequency band of $1-30\;{\rm mHz}$. In addition, controlled injection of heat signals to the suspension struts anchoring the LISA Technology Package (LTP) Core Assembly to the satellite structure allowed to experimentally estimate for the first time the phase noise contribution through thermo-elastic distortion of the LTP interferometer, the satellite's main instrument. Such contribution was found to be at $10^{-12}\,{\rm m}\, {\rm Hz}^{-1/2}$, a factor of 30 below the measured noise at the lower end of the measurement bandwidth ($1\,{\rm mHz}$).
\end{abstract}

\maketitle

\section{Introduction}
LISA Pathfinder (LPF)~\cite{lpf} will be the in-flight experiment to test key elements required for a future space-borne gravitational wave detector, as for instance the proposed eLISA mission~\cite{elisa}. LPF is an ESA mission, with some NASA contributions~\cite{drs_ref}. The main scientific instrument on-board the LPF is the LISA Technology Package (LTP), consisting of two test masses which are kept in nominal free fall by means of a precise drag-free control loop~\cite{Dolesi03}. 
A high precision interferometer~\cite{Heinzel04} measures the relative longitudinal and angular displacements with high stability. 

The main scientific goal of the mission is to reach a residual acceleration between both test masses of\, $3 \times 10^{-14} {\rm \,m/s^{2}/\sqrt{Hz}}$ in the bandwidth between $1-30\,{\rm mHz}$~\cite{temp_reqs}, as expressed in amplitude spectral density of differential acceleration fluctuations between them.
Though the sensitivity of the instrument is limited at the lower side of the band by the cross-talk actuation on the test masses and at the upper side by the interferometer sensing noise~\cite{antonucci2012}, several environmental noise sources could induce disturbances in the system that would reduce the sensitivity of the system, if they were not antecedently accounted for.

The Diagnostics Subsystem on-board LISA Pathfinder is designed precisely for that 
purpose: it will monitor magnetic fields~\cite{magnetic}, charged particles~\cite{dmu_ref} and temperature variations on the satellite~\cite{dds_pris}. It also includes a set of precision heaters and coils that will allow a characterisation of the effects in-flight and its consequent translation into noise models to be taken into account for future gravitational wave observatories in space.

The thermal case is of particular relevance since it becomes more important in the very low frequency regime~\cite{Nofrarias13}, where LISA Pathfinder has its measurement bandwidth. Also, thermal effects have been identified as a leading term in the noise contribution through different effects, namely: net forces and torques appearing on the test masses caused by asymmetric temperature distributions around them, optical pathlength variation due to temperature fluctuation of optical elements and thermo-elastic distortion of the satellite structure, including the distortion of the Optical Metrology Subsystem's optical bench itself. While there have been experiments addressing the first two ---through torsion pendulum measurements~\cite{pendulum_trento} and laser metrology set-ups~\cite{ow_paper}, respectively--- there were no empirical results providing an indication of the impact of the temperature-induced deformation of the structure hosting an instrument as the LTP, reaching picometer resolution.

Spacecraft thermal-vacuum tests are the first realistic scenario where these experiments can take place. In the following we describe the LISA Pathfinder On-Station Thermal Test (OSTT) campaign which took place late in 2011, involving for the first time thermal diagnostics experiments to be carried out during the mission and providing critical information regarding the LTP sensitivity towards thermoelastic perturbations at the LTP Core Assembly (LCA).

This test campaign was used to test the data analysis tools that are being developed for the mission. In that sense, all the data analysis was carried out with software from the LISA Technology Package Data Analysis (LTPDA) toolbox, which is a dedicated MATLAB Toolbox developed by the LPF Data Analysis team, integrating the different data analysis tools to be used during the mission operations and the posterior data analysis phase~\cite{ref_ltpda_martin}.

The paper is organized as follows. We describe the set-up during the campaign and the experiments performed in Section~\ref{ostt}. Results are presented in Section~\ref{results} together with the discussion of the distortion mechanism, following to that the thermo-elastic noise contribution to the optical metrology is presented in Section~\ref{dist_charac}, to then finally discuss our conclusions in Section~\ref{conclusions}.

\section{The LISA Pathfinder Thermal Balance and Thermal Vacuum campaign}
\label{ostt}
Thermal Balance and Thermal Vacuum test are one of the most comprehensive series of tests that a spacecraft needs to overcome before launch. The main purpose of the Thermal Vacuum test is to verify the satellite performance in its thermal design limits. This set of thermal data is required as well for correlation with the spacecraft thermal model. 
On the other side, Thermal Balance checks the performance of the spacecraft by operating all of its systems in the same simulated space conditions until thermal equilibrium is achieved.

\begin{figure}[t!]
\begin{center}
\includegraphics[trim=0cm 0cm 0cm 12cm, clip=true,width=7cm]{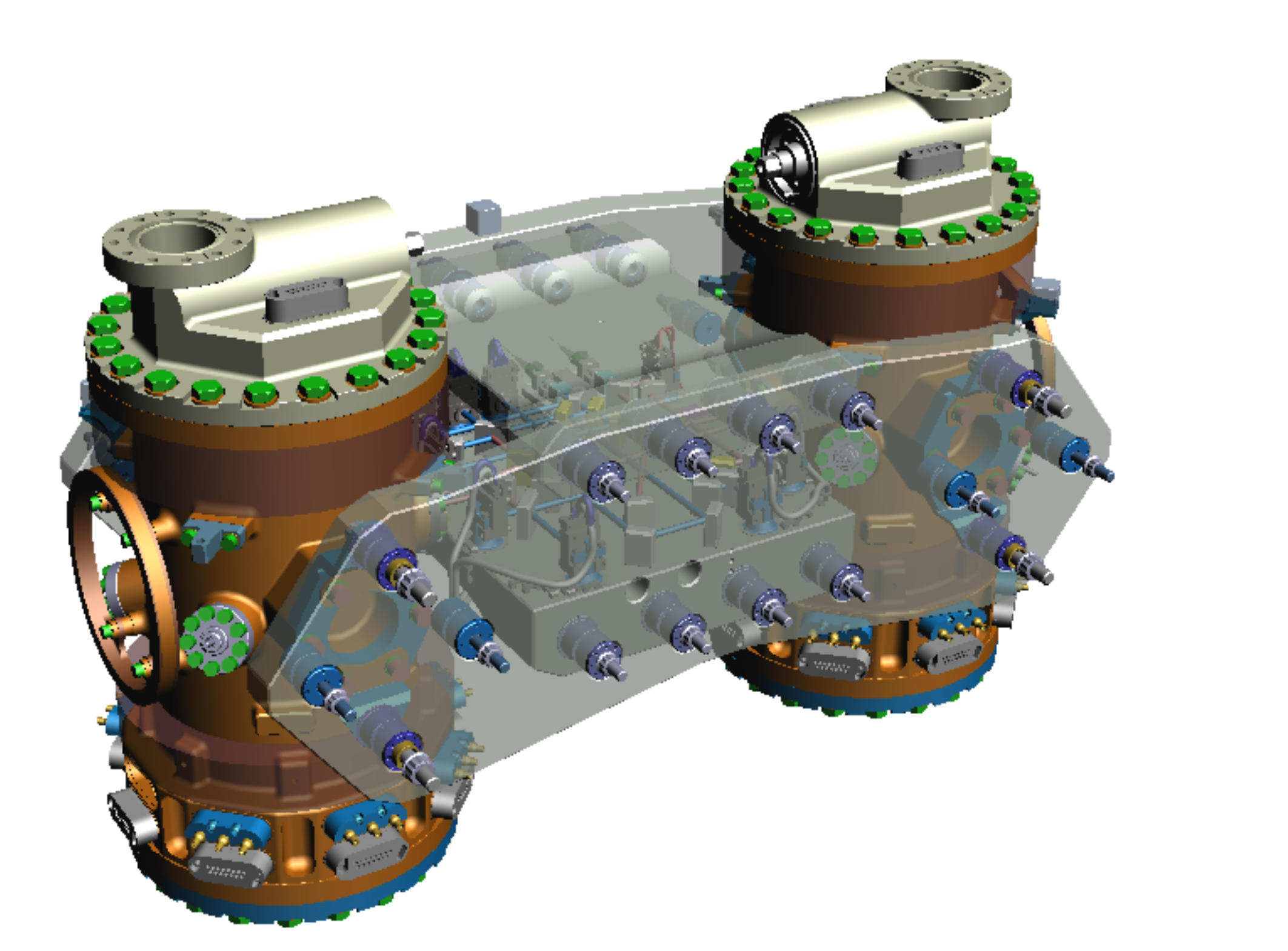}
\includegraphics[width=7cm]{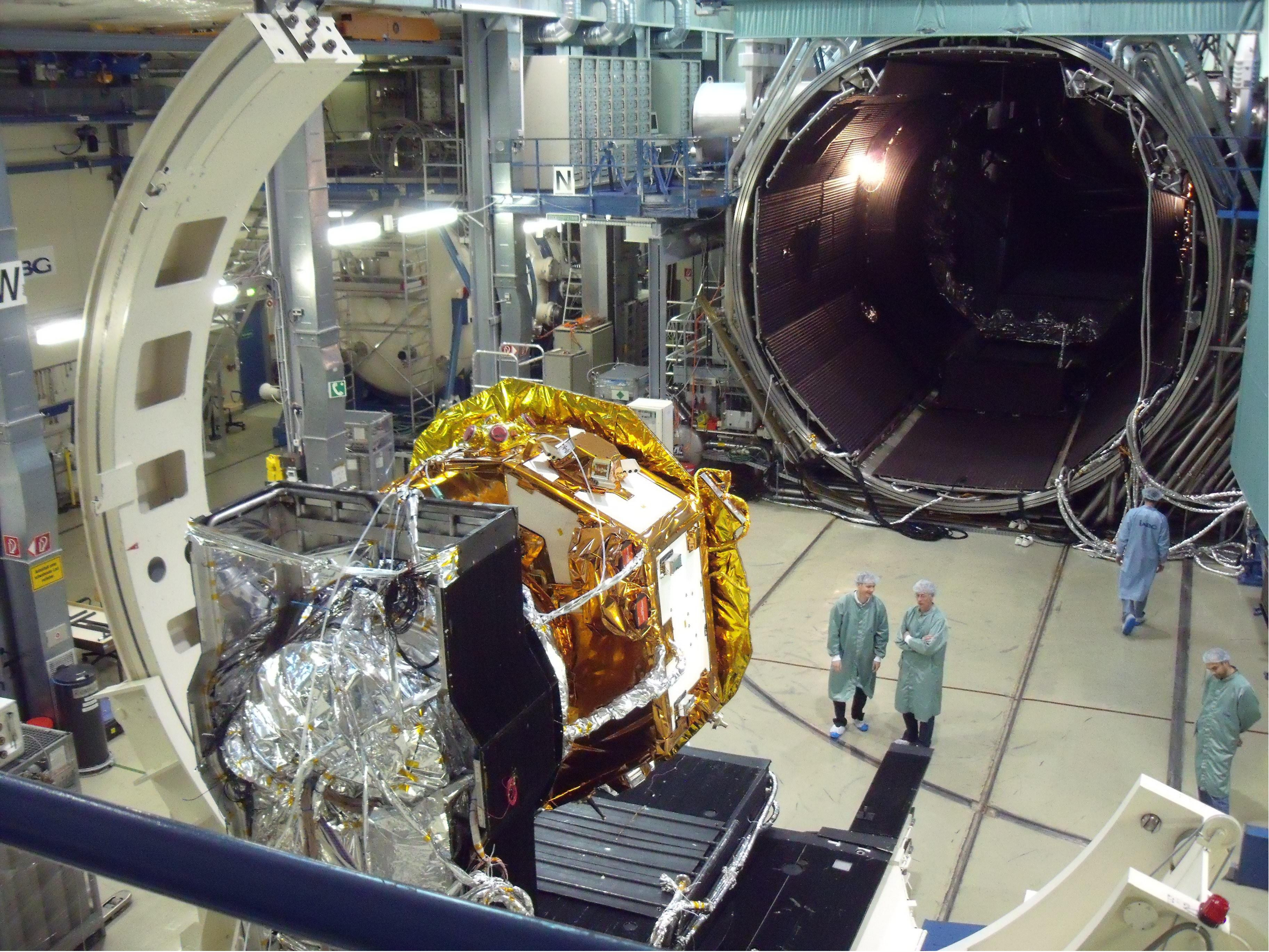}
\caption{{\it Left:} General view of the LTP Core Assembly (LCA), with both Vacuum Enclosures that host the Inertial Sensors. The optical bench is located between them. Image credits: Airbus Defence and Space, Germany. {\it Right:} IABG mbH space vacuum chamber with the LPF during the OSTT campaign. Image credits: Airbus Defence and Space UK.}
\label{ostt_picture}
\end{center}
\end{figure}

For LISA Pathfinder, these represented an extensive campaign of tests conducted at the IABG mbH space simulator, in Ottobrun (Germany) in the framework of the On-Station Thermal Test campaign --see Figure~\ref{ostt_picture}. During the campaign, the spacecraft was operated in space conditions, i.e. with a nominal pressure below $10^{-4}$\,Pa and powered up by means of a sun simulator, consisting of an array of high power lamps mimicking the radiation that will hit the spacecraft in orbit.

\subsection{Subsystems in the LISA Pathfinder Core Assembly}

The instrumentation on-board the satellite was almost the complete flight version of the LPF satellite. Most parts of the different satellite units were already integrated and operative. Here we provide a brief description of the main subsystems in the LCA and comment any modifications to the setup with respect the flight version.

\paragraph{Gravitational Reference Sensor}

During the Thermal Balance and Thermal Vacuum campaign, the Electrode Housings and the test masses of the Inertial Sensors were not present and were replaced by piezoelectric driven mirrors acting as end-mirrors of the interferometer~\cite{felipe}.

\paragraph{Optical Metrology Subsystem}

The Optical Metrology Subsystem is composed by the Reference Laser Unit, the Laser Modulator, the Phasemeter~\cite{Heinzel04}, the Optical Bench~\cite{robertson_cqg_2013} and the Data Management Unit (DMU)~\cite{dmu_ref}, all of them integrated in the spacecraft during the campaign. In the Optical Bench four different interferometer measurements are taking place. The main two interferometers are the $\rm x_1$ that measures the distance between Test Mass 1 and the optical bench and the $\rm x_{12}$ provides the relative distance between the two test masses
 

\begin{figure}[t]
\begin{center}
\includegraphics[width=12cm]{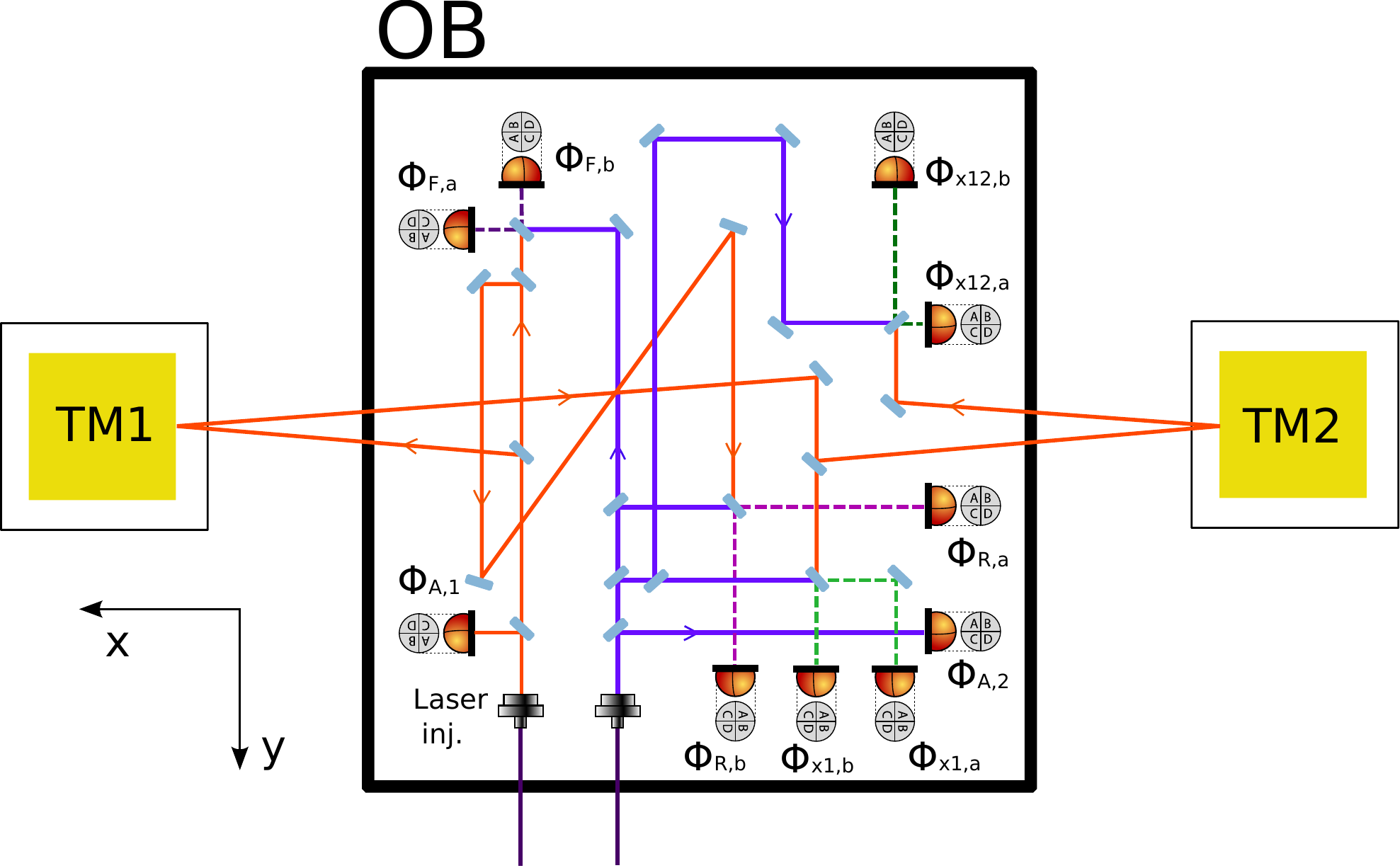}
\caption{Schematic of the interferometer main measurements: ${\rm \Phi_{12}}$ quadrant photo-diodes measure relative displacement of Test Mass 1 with respect to Test Mass 2, and angular relative position on Y and Z axes between them. On the other hand, ${\rm \Phi_{1}}$ quadrant photo-diodes measure the same displacements and angles on Test Mass 1 but with respect to the spacecraft. The final relative displacement measurements ${\rm x_{12}}$ and ${\rm x_{1}}$ are obtained by subtracting the reference measurement ${\rm \Phi_{R}}$ to both ${\rm \Phi_{12}}$ and ${\rm \Phi_{1}}$ respectively. The suffixes {\it a, b} refer to each of the redundant photodiodes of each interferometer measurement, and the dashed lines represent the recombined beams for each measurement. At the OSTT campaign the Test Masses were replaced by mirrors.}
\label{ob_measurements}
\end{center}
\end{figure}

A third interferometer, the {\it reference} interferometer is sensitive to environment perturbations and pathlength noise originated out of the ultra-stable Zerodur optical bench. This read-out is subtracted from the rest of the interferometers read-outs in order to remove common-mode disturbances and increase their
sensitivity. In addition to these three interferometers, the optical bench is also equipped with a {\it frequency} interferometer similar to the {\it reference} interferometer but with an intended 38\,cm pathlength mismatch which makes it specifically sensitive to laser frequency fluctuations. This readout acts as a control signal of the Reference Laser Unit. 

The photodiodes used for these measurements are actually quadrant photodiodes, therefore they are also sensitive to displacements and angular fluctuations of the beams. Such feature is used to measure the test masses attitude around Y ($\eta_1$, $\eta_2$) and Z ($\phi_1$, $\phi_2$). Two different approaches are used to measure these angles~\cite{Heinzel04, angles_meas}:
\begin{itemize}
\item {\it Differential Power Sensing} (DPS) measurements, where beam displacements are measured on the quadrants and test mass angles inferred. Strictly, such measurement provides the average displacement of the two recombined beams on each quadrant photodiode.
\item {\it Differential Wavefront Sensing} (DWS) measurements, where the test mass angles are obtained by measuring the relative angles between the two beams on each quadrant photodiode. 
\end{itemize}

While the DPS technique has a wider dynamic range, the DWS is used as a measurement signal for the different control loops because of its better sensitivity. 

For redundancy issues, each interferometer's measurement is performed by actually two quadrant photodiodes. Considering that there are as well two additional photodiodes	to control the amplitude stability of each injected laser beam \footnote{The amplitude detectors are single element photodiodes but, due to the lack of availability of convenient space-qualified single element photodiodes they are implemented as one element of a quadrant photodiode.}, a total of ten quadrant photodiodes are placed on the LTP optical bench -- see Figure~\ref{ob_measurements}.

\paragraph{Temperature Diagnostics}

\begin{figure}[t!]
\begin{center}
\includegraphics[width=12cm]{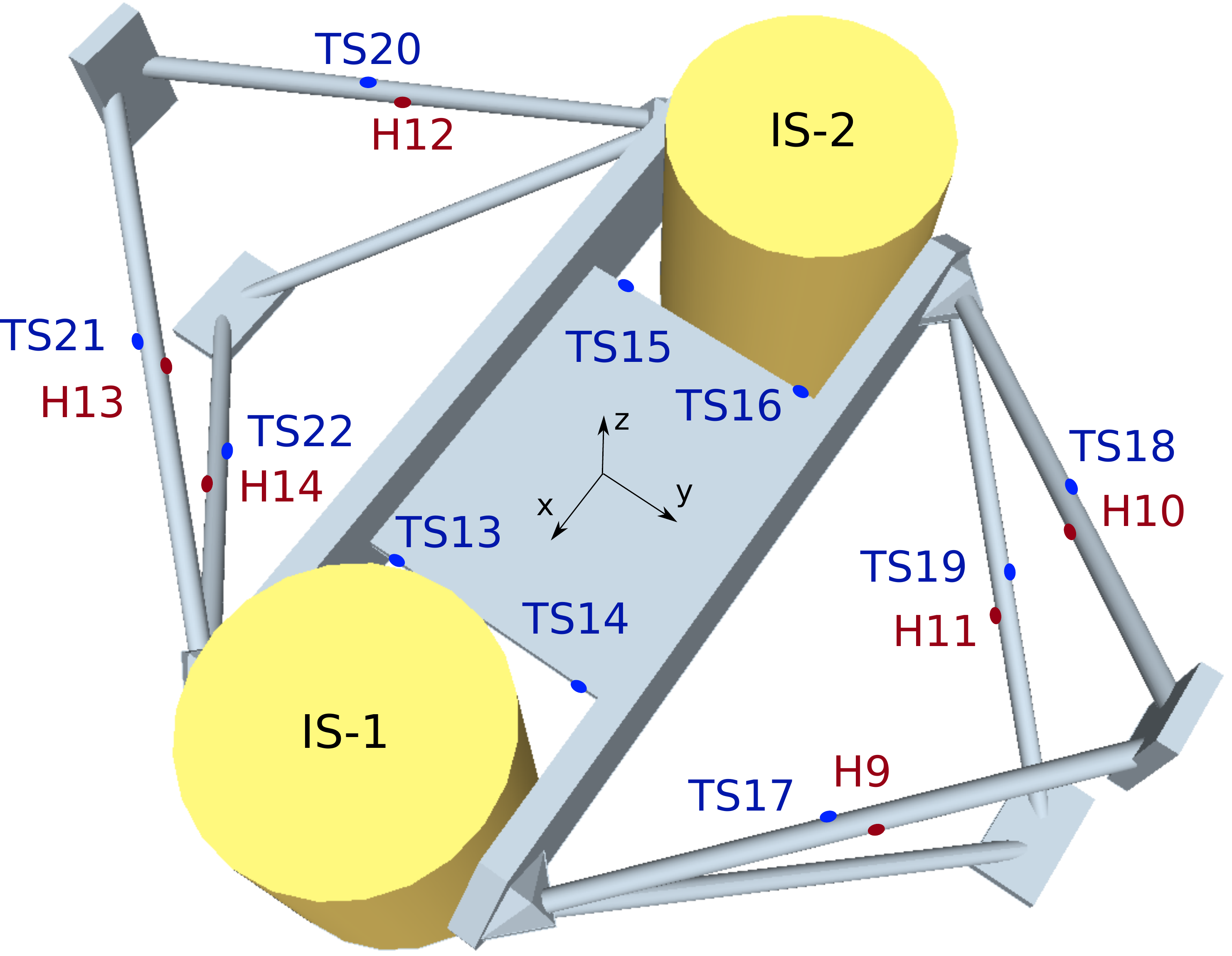}
\caption{Schematic of the LCA and its series of diagnostics heaters and temperature sensors on the different Suspension Struts and on the optical bench, as during the OSTT campaign.}
\label{strutslayout}
\end{center}
\end{figure}

The thermal part of the Diagnostics Subsystem in LPF~\cite{dds_pris} is composed by a set of 24 sensors and 14 heaters distributed through the LCA. The fact that the Inertial Sensor was not part of the campaign implied a 
relocation of some of the temperature items. Regarding the temperature sensors, 8 thermistors were relocated around the dummy Inertial Sensor Housing structures and 6 placed on the inner sides of the LCA support structure, leaving at their flight representative locations only the 6 Suspension Struts temperature sensors and the 4 sensors located on the optical bench  ---see Figure~\ref{strutslayout}. Despite of the modifications, the temperature measurement system (TMS) could perform correctly. 
With respect to the heaters, only the 6 Suspension Strut heaters were present and placed in their design locations. As a consequence, among the different heating experiments planned for the mission, only the thermistors and heaters concerning the Suspension Strut heating were available for test.

\subsection{Thermal experiments during the campaign}
\label{heater_seq}

The campaign was divided in two main stages, where experiments involving many different subsystems of the spacecraft were repeated at two different temperatures, i.e. the {\it hot case} at $30.5^\circ\, {\rm C} \pm 0.5^\circ\,{\rm C}$ and the {\it cold case} at $9.5^\circ\, {\rm C} \pm 0.5^\circ\,{\rm C}$, reaching temperatures of around $26^\circ\,{\rm C}$ and $12^\circ\,{\rm C}$ respectively in the LCA~\cite{felipe}. In both of them a series of experiments to test system performance and to characterise different spacecraft subsystems were run and, as regards of the thermal diagnostics, the execution of different telecommand sequences of heater activations was included. These experiments were carried out during the {\it cold} case and were planned as follows:
\begin{itemize}
\item Phase 1: Continuous heat injection in Heaters 9 and 11 (14h).
\item Phase 2: Individual heater activations in series of pulses to all strut heaters (6 $\times$ 2h). 
\item Phase 3: Combined heater activation for thermoelastic stress tests (10h).
\item Phase 4: Relaxation time (12h).
\end{itemize}

{\it Phase 1} experiment was aimed to test long-term heating system performance while {\it Phase 2} experiments were providing most of the information regarding the system response to local ---at a single {\it Suspension Strut level}--- temperature fluctuation. The incidence of such experiments to the interferometer readout was expected to be linear, what was checked with data from {\it Phase 3}. {\it Phase 4} provided with a set of noise data that will be used to analyse the temperature noise contribution during normal performance as well as with information regarding the relaxation times needed to recover from heater-induced conditions.


The whole data communication chain between the spacecraft and the operations center was already through the flight model DMU and OBC (On-Board Computer), and the spacecraft kept being powered up from the solar-like radiation energy provided by an array of high power lamps in the vacuum chamber.

\section{Thermo-optical characterisation of the LISA Pathfinder Core Assemby}
\label{results}
\begin{figure}[t!]
\begin{center}
\includegraphics[width=5.45cm,angle = -90]{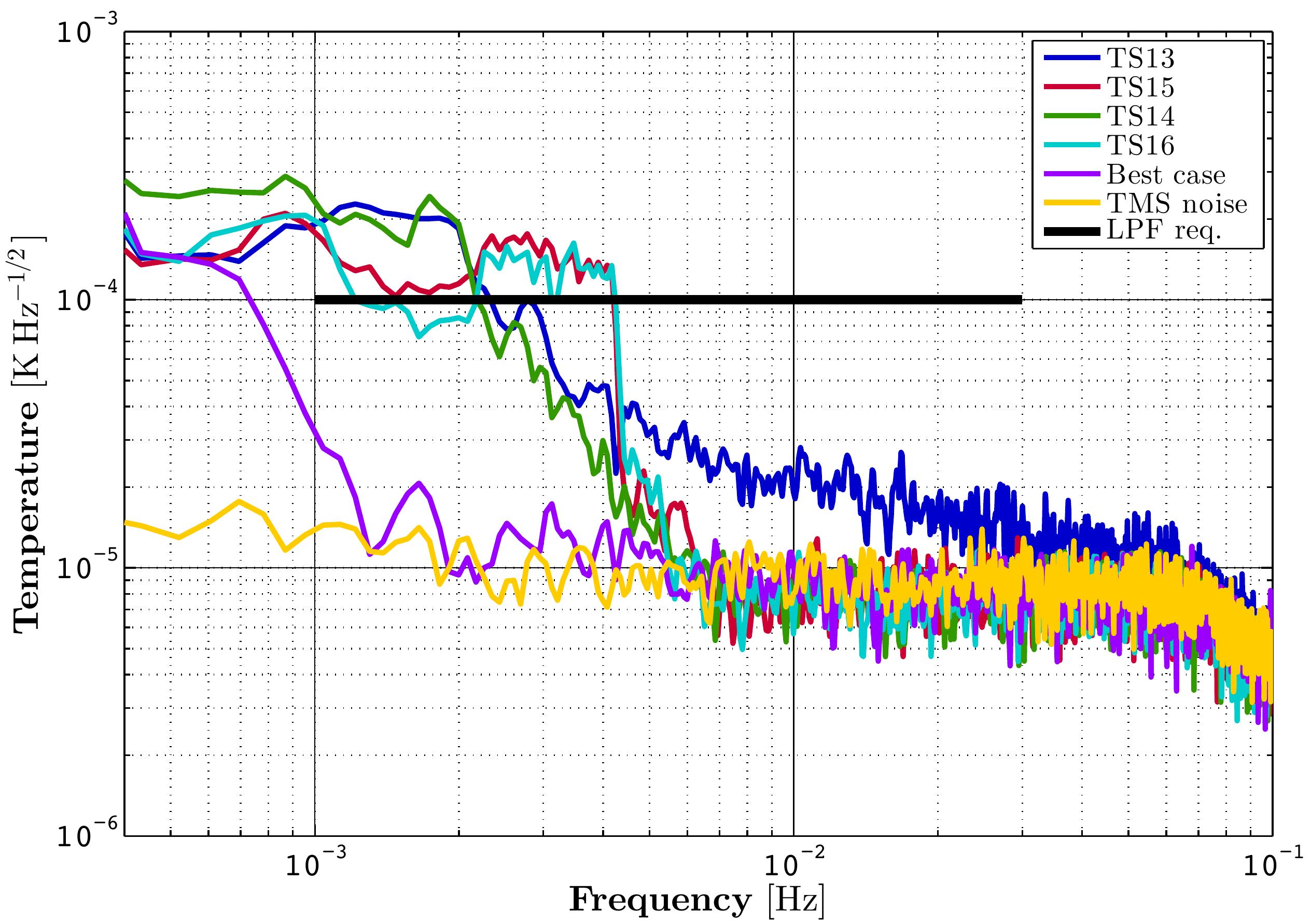}
\includegraphics[width=5.45cm,angle = -90]{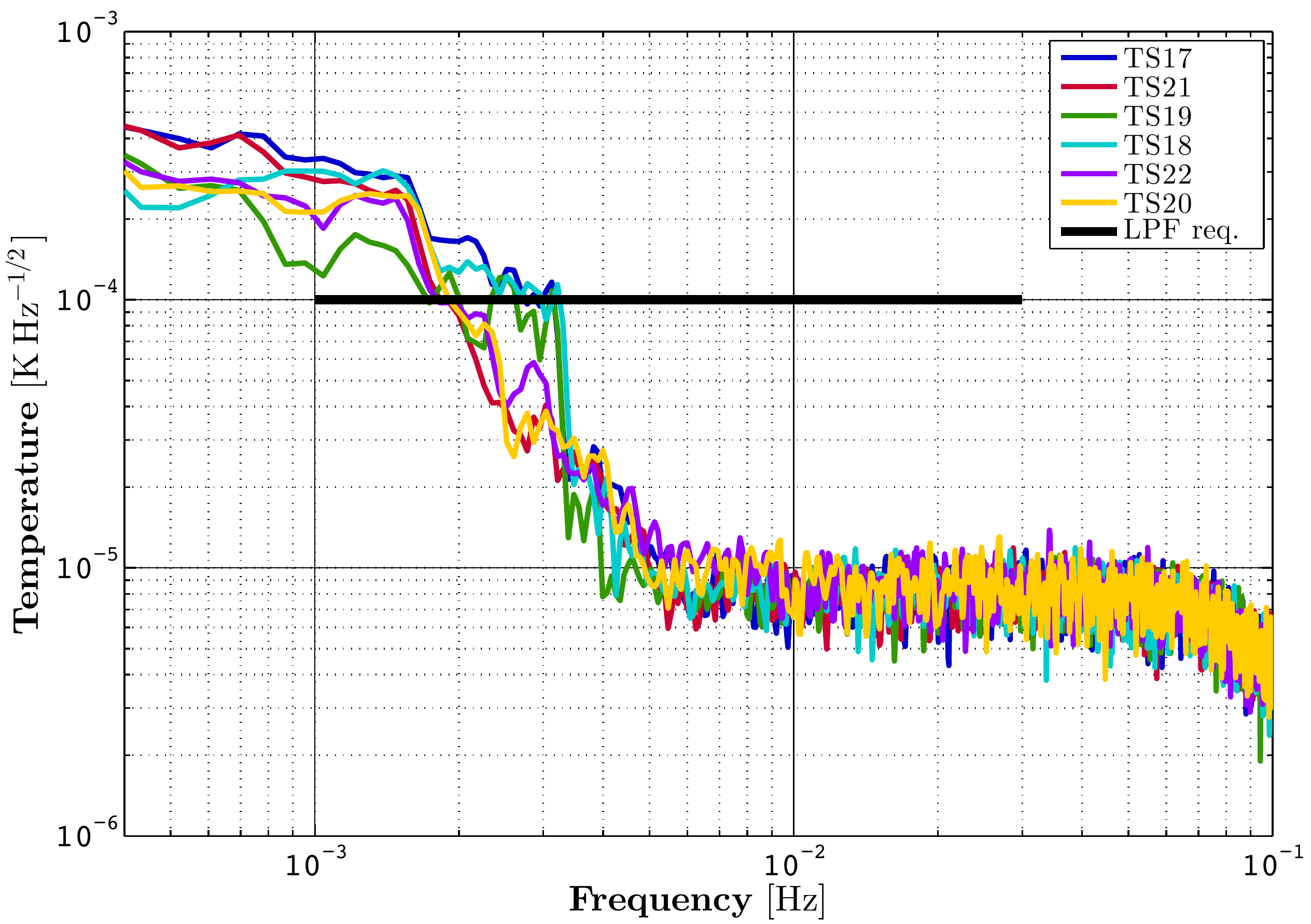}
\caption{{ \it Left:} Power Spectral Density (PSD) for the optical bench temperature sensors, together with the best performance case of the temperature measurement subsystem during the campaign (labelled as {\it Best case}), obtained outside the LCA by a {\it relocated} sensor. The system reference noise (labelled as {\it TMS noise}) sets the floor for the temperature measurement. All these spectra correspond to temperature measurements taken some days before the thermal experiments, during the {\it hot} case of the campaign. {\it Right:} PSD of the different Suspension Strut temperature sensors for the same times. At both Figures, the sudden noise falls observed for instance on TS15-19 are attributed to non-linear effects of the ADC, caused by the presence of too high temperature drifts during the campaign --see~\ref{adcerror}.}
\label{ifoTP4}
\end{center}
\end{figure}

\subsection{Temperature environment characterisation}
\label{data_conditioning}
The temperature measurement system kept active through all the tests of the campaign, allowing a complete characterisation of the environment whilst the different planned experiments were taking place.
Before entering in the analysis of the measurements obtained through the set of temperature sensors, it is worth describing the temperature acquisition readout and the associated data pre-processing. 

The LPF temperature measurement system is based on series of Wheatstone bridges centred at different temperatures in order to perform all the measurements close to the linearity of the system, covering a total temperature range between around $8 - 30$ degrees Celsius~\cite{tms_ref}. The data is onboard measured at $0.83\,{\rm Hz}$ and --due to limitation on the downlink telemetry budget for diagnostics-- filtered with a Butterworth filter and then downsampled to $0.21\,{\rm Hz}$. Since the filter is applied directly to the measured data, each change of scale is equivalent to a changing signal bias for the filter, what induces a transient response to the downsampled data. As a consequence of the previous, spikes can appear along the data at each change of scale. In order to clean the signal from such spikes, sets of samples are removed and interpolated, while the remaining offsets are carefully identified and subtracted. 

\label{tms}
The best temperature stability during the OSTT campaign was found on a {\it relocated} sensor outside the LCA (case of TS9, labelled {\it Best case} in Figure~\ref{ifoTP4}), achieving a noise level of $10^{-5}{\rm K}\,{\rm Hz}^{-1/2}$ through most frequency band, in practice achieving the electronic noise floor as expressed by the {\it reference} temperature measurements, i.e. measurements performed with high stability resistors instead of thermistors (labelled {TMS noise} in Figure~\ref{ifoTP4}). 

Such low levels were not achieved inside the LCA because of a high temperature drift that induced a feature in the spectrum of temperature fluctuations due to a cyclic error associated to analog-to-digital converter (ADC) nonlinearity~\cite{pep_adc}. Such an error is related with imperfections at the quantization levels of the ADC. \ref{adcerror} focuses on this particular issue, and how the error coming from the thermal drift is identified. 

Moreover, slightly different temperature noise levels at different sides of the optical bench caused sensors TS15 and TS16 to manifest higher noise levels than TS13 and TS14 though they had similar temperature drifts --see the bump affecting only TS15 and TS16 between $2\,{\rm mHz}$ and $4\,{\rm mHz}$ at Figure~\ref{ifoTP4}, {\it left}. This can be explained by the attenuation effect of the ADC non-linear error higher harmonics when the input signal combines a drift with certain levels of noise~\cite{pep_adc}. A higher temperature noise level on the +X side --where TS13 and TS14 are located-- can actually reduce the effect of this error, as it happens in Figure~\ref{ifoTP4} to TS15 and TS16. The same effect appears on the temperature stability plots of the struts (Figure~\ref{ifoTP4}, {\it right}), where a lower noise profile on the +Y side of the LCA between $2\,{\rm mHz}$ and $3\,{\rm mHz}$ induces higher ADC non-linear while on the -Y side the error is dumped.

The temperature drift causing the different bumps and sudden falls in Figure~\ref{ifoTP4} appeared due to test schedule limitations, i.e. not allowing the instrument to achieve a complete steady state. The real drift during flight operations is thus expected to be smaller. 


\subsection{Response of the $\rm x_{1}$ and $\rm x_{12}$ interferometers to heat inputs}

\begin{figure}[t]
\begin{center}
\includegraphics[width=5.4cm,angle = -90]{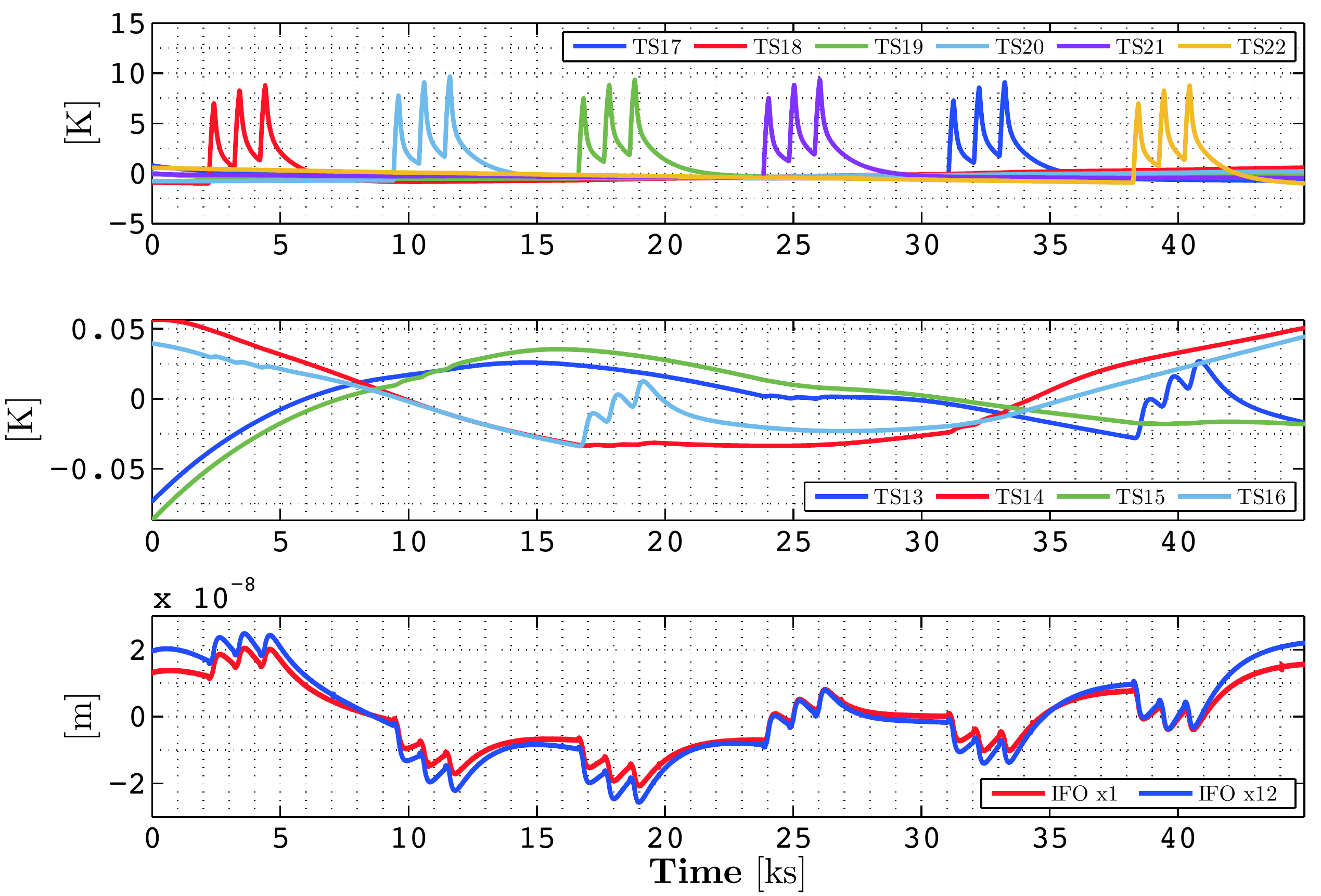}
\includegraphics[width=5.4cm,angle = -90]{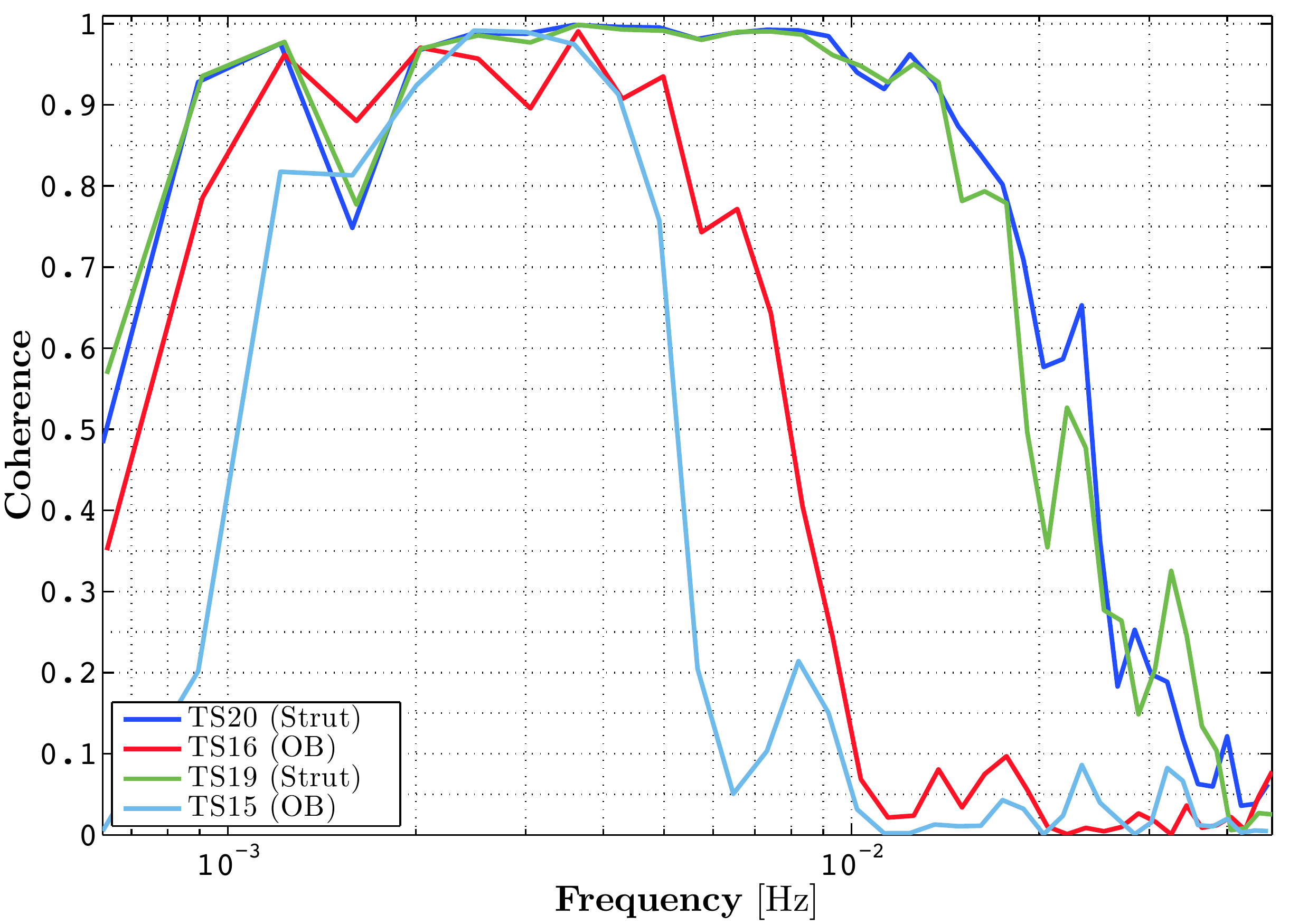}
\caption{ On the {\it left} side, the temperature measurements at the different struts ({\it upper} plot), the optical bench temperatures ({\it centre} plot) and  the displacements measured by the interferometer (IFO, {\it lower} plot), all of them detrended with a 1st-order polynomial. On the {\it right}: coherences between different strut sensors (TS19 and TS20) and interferometer's channel ${\rm x_{12}}$ together with the coherences from their closest optical bench (OB) sensors (TS15 and TS16 respectively) to the same interferometer measurement.}
\label{heaters_run}
\end{center}
\end{figure}

During the {\it Phase 2} of the diagnostics experiments, a series of three pulses of $200\,{\rm s}$ in periods of $1000\,{\rm s}$ and a power of $2\,{\rm W}$ were applied individually to each heater, producing temperature increments in the respective struts around $8\,{\rm K}$ per pulse and inducing immediate observable consequences on the interferometer channels with amplitudes around $\pm 10\,{\rm nm}$, as shown in Figure~\ref{heaters_run}. 
In order to better distinguish the effect, a 1st-order detrend is applied to all time series in the Figure~\ref{heaters_run}.

Temperature increments at the optical bench clearly indicate that temperature variations in the lower struts (TS19, TS22) have a thermal impact on the optical bench much larger than the upper struts (TS17, TS18, TS20, TS21). Indeed, the central panel in Figure~\ref{heaters_run} shows how the temperature sensors in the optical bench have a clear response for the former and a negligible reaction ---out of the environmental drift--- for the latter. 

The difference in the heat conduction through the parts between the upper and the lower struts to their optical bench closest sensor cannot justify such a significant difference of more than one order of magnitude (about a factor 13). On the other hand, none of the struts has visibility to the sensors which could create a direct radiative link.

An explanation of this effect must be sought in the fact that, opposite to the upper (+Z) struts --see Figure~\ref{strutslayout}--, the screws attaching the lower (-Z) struts to the optical bench assembly present direct visibility to their closest optical bench sensor, establishing a direct thermal radiative link. If we consider on one hand the screw temperature being increased at least up to $2\,{\rm K}$ through thermal conduction, a heated area of at least $20-22\,{\rm mm^2}$ and a screw emissivity of around $0.5$ and, on the other hand, a sensor area of around $10\,{\rm mm^2}$ with high absorption ($\sim 0.9$) and a thermal resistance to the optical bench of $\sim 100\,{\rm K/W}$, increments of $\sim 10\,{\rm mK}$ should be expected in these  optical bench sensors. This is already about one order of magnitude above the equivalent case with just conduction ---which is the case for the higher struts.

Nevertheless, the lower left hand plot in Figure~\ref{heaters_run} shows how the interferometer perturbation repeats the same pattern of signals for all the heater activations, regardless of the above discussed asymmetry on +Z and -Z struts. In order to further investigate this effect, we computed the coherence function between the interferometer phase readout and the four temperature sensors involved in the -Z struts experiments. 

\begin{figure}[t]
\begin{center}
\includegraphics[width=5.35cm,angle = -90]{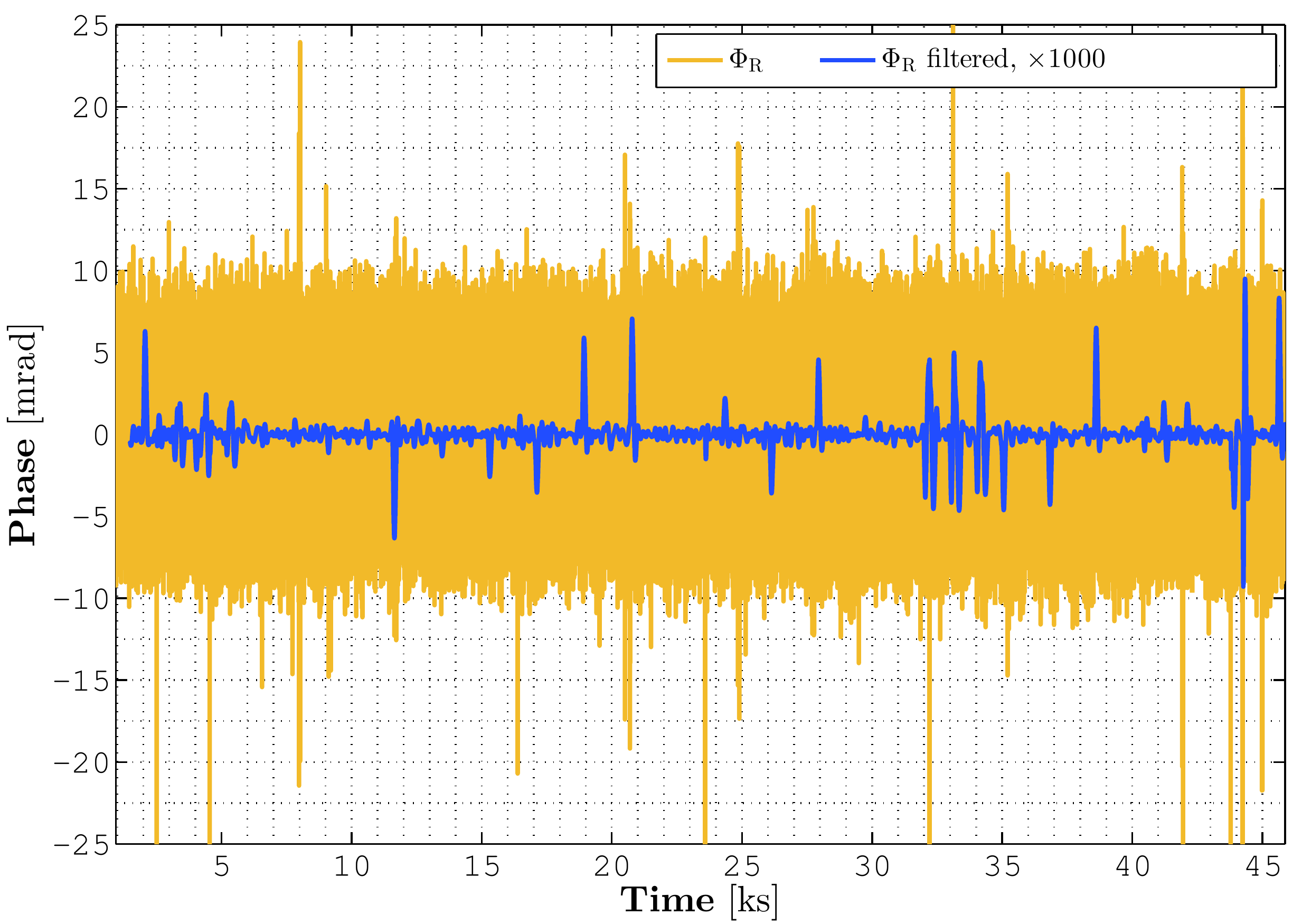}
\includegraphics[width=5.35cm,angle = -90]{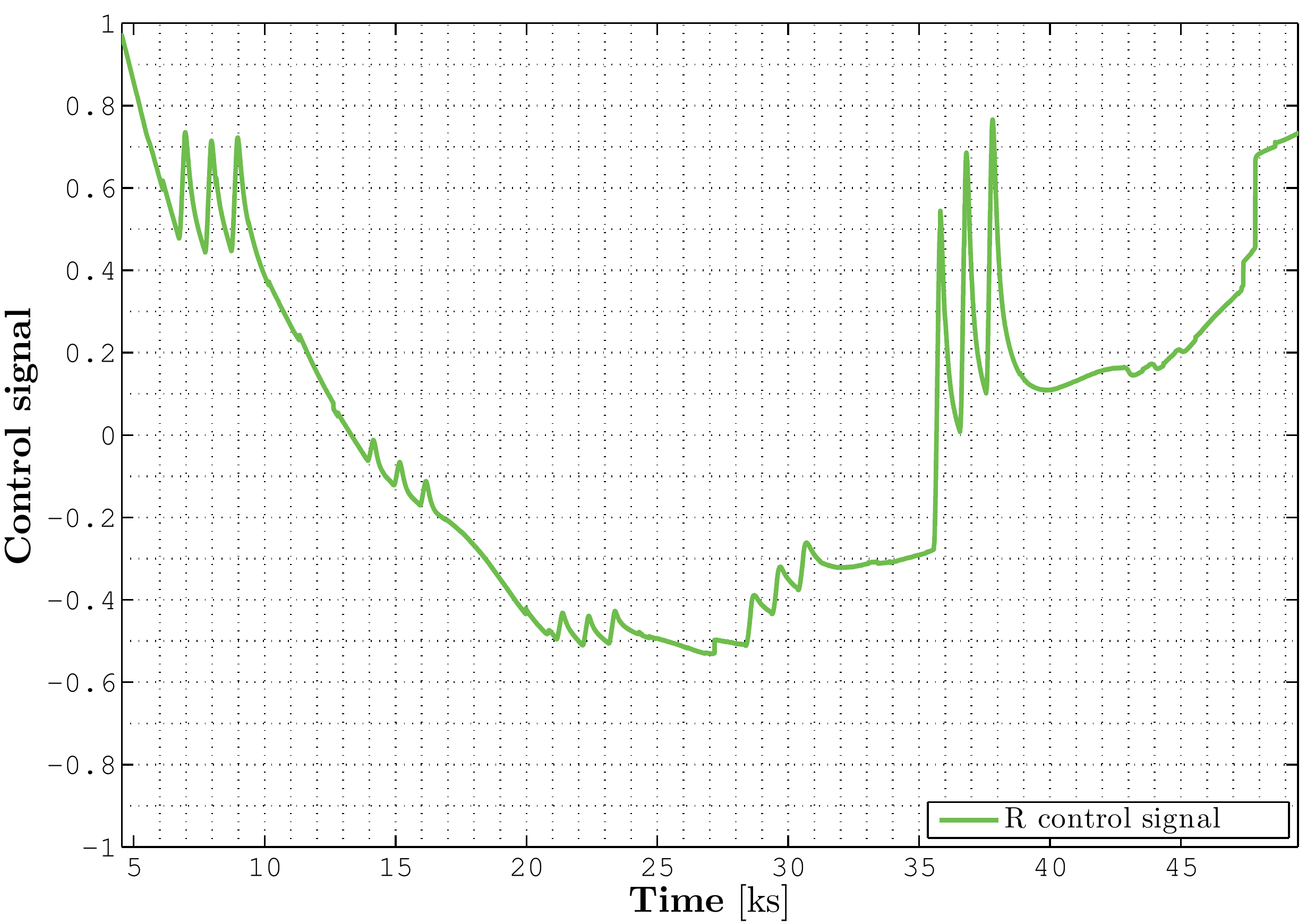}
\caption{{\it Left:} Readouts of the reference interferometer $\Phi_{\rm R}$ and the same measurement filtered with a 4th order lowpass filter with a cut-off frequency of $5\,{\rm mHz}$. {\it Right:} OPD control signal applied to the piezo actuator that controls the differential optical pathlengh fluctuations.}
\label{R_ifo_data}
\end{center}
\end{figure}

As shown in Figure~\ref{heaters_run} ({\it right}), the analysis confirms that the signal observed in the interferometer is not coherent with temperature variations observed in the optical bench ($\simeq 8$\% coherence 10\,mHz with sensors TS15 and TS16) and is strongly correlated with the temperature increase in the struts ($\simeq 92$\% coherence at 10\,mHz with sensors TS19 and TS20). This suggests that, though the Zerodur plate is being heated up through radiative effects when activating lower-strut heaters, the interferometer sensitivity from this effect is negligible in comparison to the general strut heating effect, discarding the Zerodur plate heating up as the mechanism describing the main perturbation and pointing to a global elastic distortion effect on the LCA.

\subsection{Response of the static interferometers to heat inputs}

The fact that the longitudinal motion signals ${\rm x_{12}}$ and ${\rm x_{1}}$ present similar amplitudes no matter which strut is being heated (though the ${\rm \Phi_{12}}$ optical path is a factor 2 longer than the ${\rm \Phi_{1}}$) suggests that the distortion mechanism does not involve the vacuum enclosures with the inertial sensors and must be sought on the optical bench itself.

In order to further investigate the origin of the response of the main interferometer channels --$\rm x_{1}$ and $\rm x_{12}$--, we explore the readouts of the two remaining interferometers. The frequency and reference interferometers are {\it static} interferometers in the sense that, apart from the fibres, they do not have moving parts outside the bench.

\begin{figure}[t¡]
\begin{center}
\includegraphics[width=5.35cm,angle = -90]{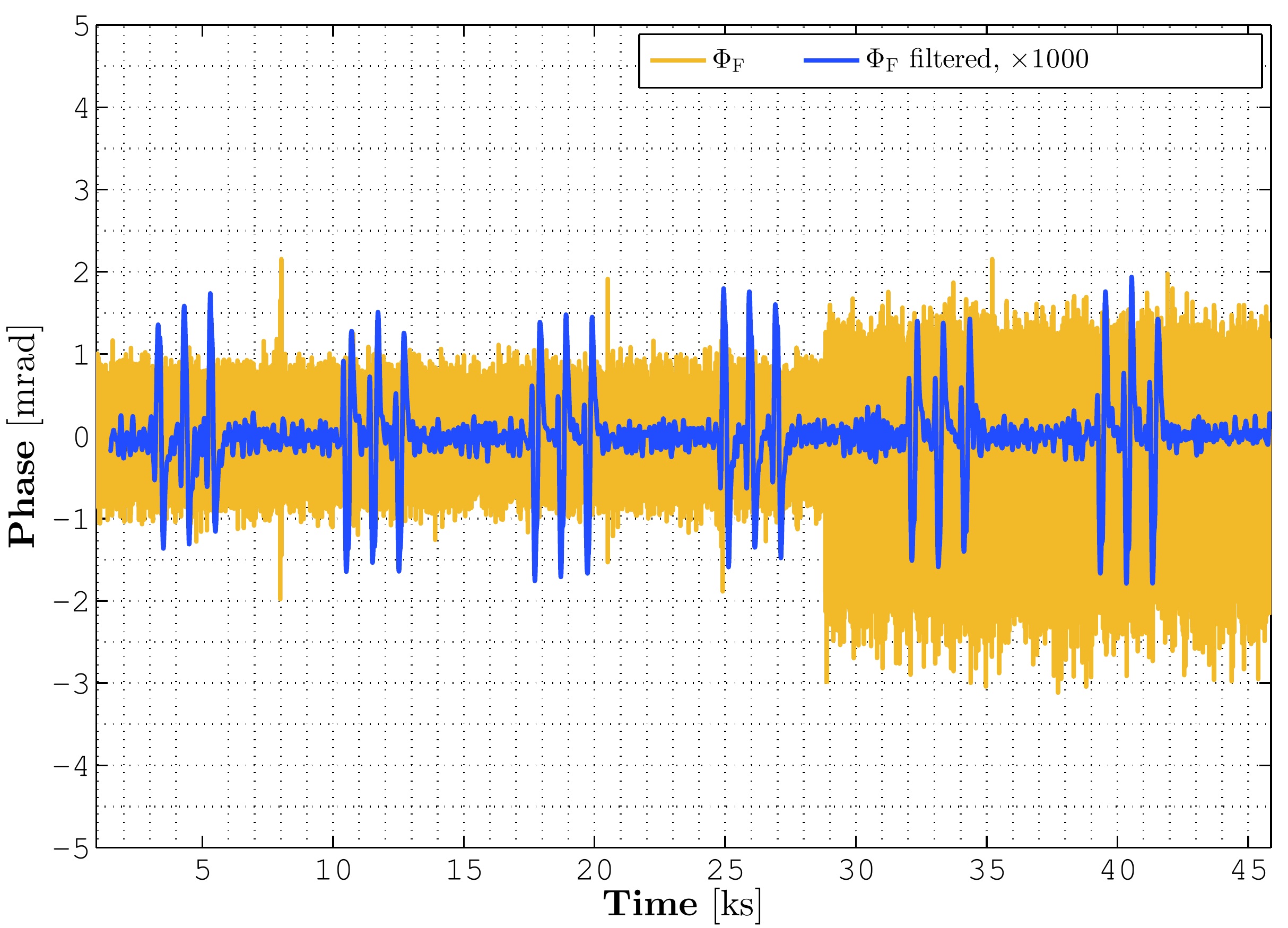}
\includegraphics[width=5.35cm,angle = -90]{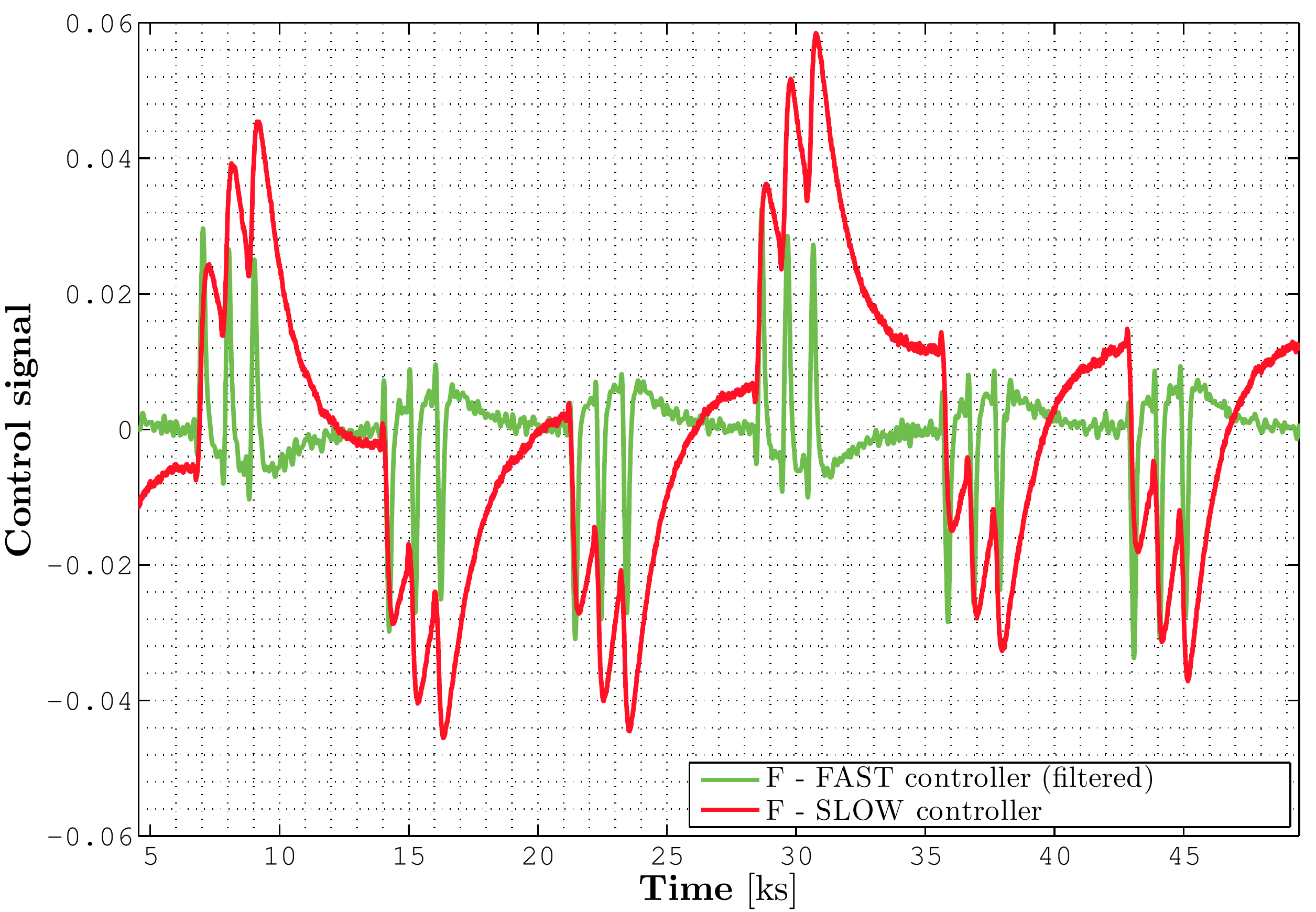}
\caption{{\it Left:} Readout of the frequency interferometer $\Phi_{\rm F}$ and the same measurement filtered with a 4th order lowpass filter with a cut-off frequency of $5\,{\rm mHz}$. {\it Right:} Control signals applied to the actuators for frequency noise stability. The fast controller --filtered here with the same filter than in the adjacent plot-- is related to a piezoelectric transducer acting on the laser head while the slow control signal is related to a temperature control loop on the laser head as well. The slow control signal shows how a similar effect is affecting the optical bench no matter the heater being activated.}
\label{F_ifo_data}
\end{center}
\end{figure}

The reference interferometer readout is roughly the phase measured at $\Phi_R$ and includes mainly all the optical pathlength fluctuations outside the optical bench --excepting eventual test masses motion. In addition, it is used to actively stabilize the optical pathlength difference (OPD) between fibres~\cite{hechenblak} and, since its information may be shaped by the noise its relevant information is found on its control signal rather than on the signal itself. In Figure~\ref{R_ifo_data}, the OPD control signal shows how the reference interferometer is sensitive to each of the heat injection series applied. As observed, the activation of Heater 9 (close to TS17) produces the largest response, followed by the activation of Heater 10 and, at third position, the activation of Heater 19, all of them at the +Y side of the LCA. However, the disturbance is significantly removed from the final reference interferometer readout and only fast optical pathlength corrections are observed, specifically ones from H9 activation. 
Due to the vicinity of this heater to the optical fibres, this suggests that a possible interaction with the optical fibres could cause this effect. However, the eventual consequences of this effect are cancelled in the common-mode rejection and do not explain the observations in Figure~\ref{heaters_run}.

Regarding the other {\it static} interferometer, the frequency interferometer measurement is used to control the frequency fluctuations in two different ways: a piezoelectric transducer to stabilize {\it high} frequency fluctuations and a thermal control loop to regulate the low frequency oscillations by controlling the temperature at the laser head, being the {\it fast} controller nested inside the loop of the {\it slow} controller~\cite{hechenblak}. Since the final readout is the result of subtracting $\Phi_R$ to $\Phi_F$ this interferometer is expected to present significantly less noise than the reference one, and actually a reduction factor of 10 is found between them --see Figure~\ref{F_ifo_data}, {\it left}. After removing the reference measurement the effect is much more homogeneous and, as expected, the proximity to the optical fibres no longer explain the effect. Regarding the control signals in Figure~\ref{F_ifo_data}, {\it right}, quick kicks corresponding to corrections from the fast controller (obtained after low-pass filtering the whole signal) are combined with a quite smoother correction by the other controller. The signal amplitudes of the {\it slow} controller present approximately the same amplitude no matter the heater being injected, which is coherent with the observations in the main interferometers. Since the frequency interferometer can only be reporting disturbances from the optical bench, it is quite evident from Figure~\ref{F_ifo_data} that the optical bench must be being distorted here following a unique deformation mode.

\begin{figure}[t]
\begin{center}
\includegraphics[width=5.35cm,angle = -90]{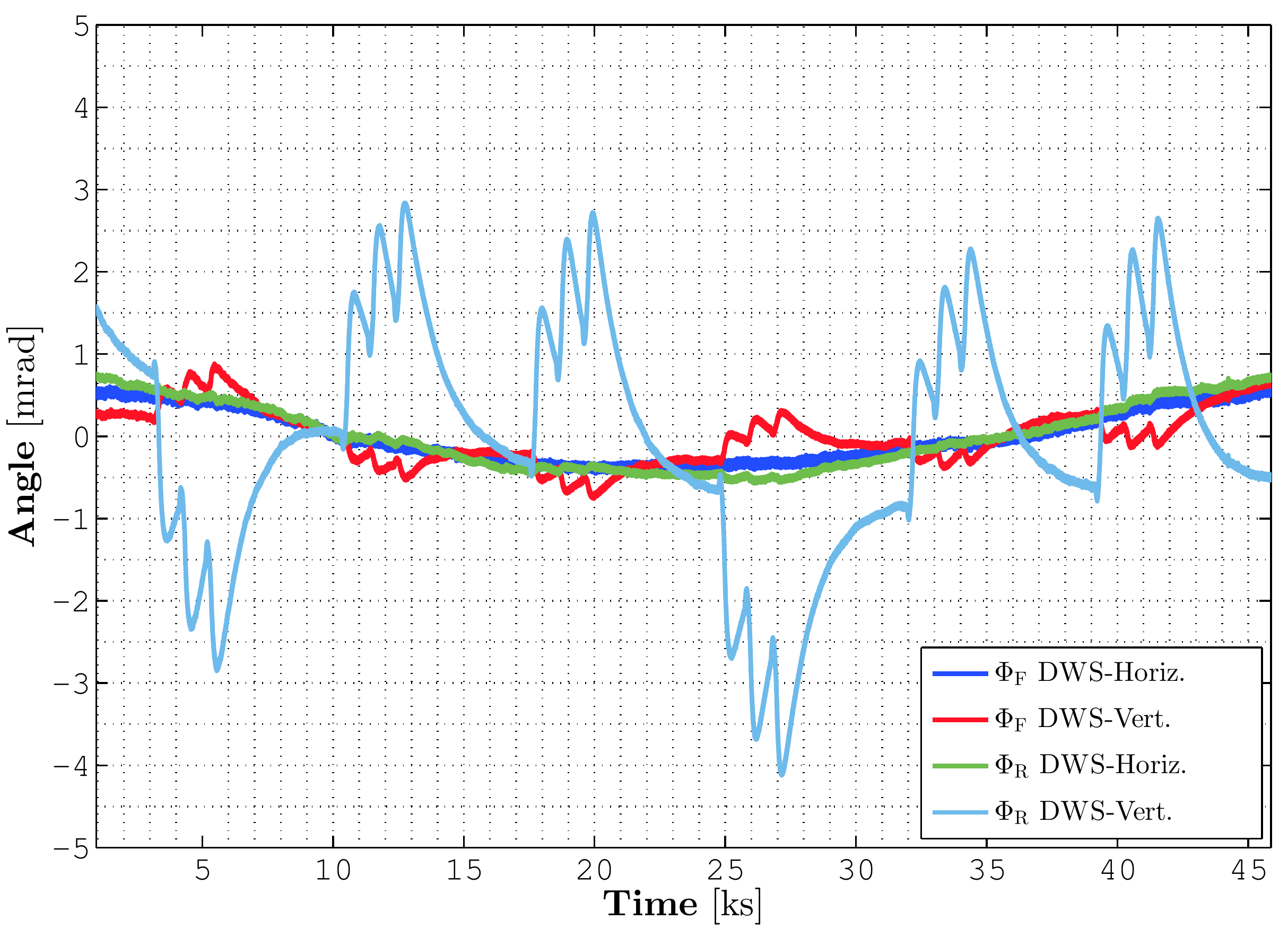}
\caption{DWS angles of the vertical and horizontal incident beams on the reference and frequency interferometers, after a 1st order detrend.}
\label{RF_DWS_angles}
\end{center}
\end{figure}

\begin{figure}[h!]
\begin{center}
\includegraphics[width=7cm]{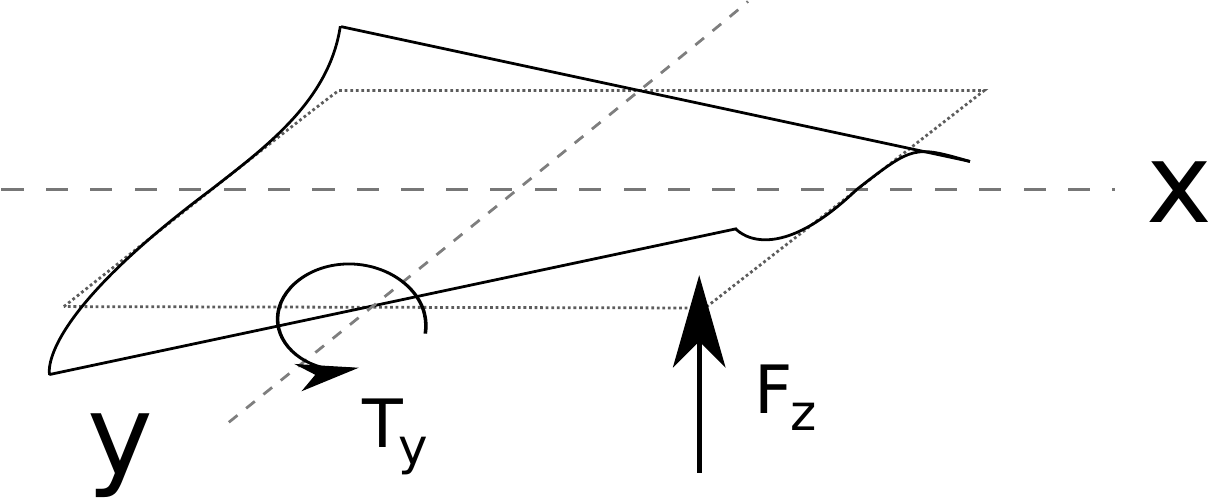}
\caption{Schematic of the torsion mechanism on Y produced by the vertical component of the force exerted by a heated strut. The case of the image would correspond to a lower strut heater activation such H11 or H14. This distortion mechanism is coherent with the observations.}
\label{ob_torsion}
\end{center}
\end{figure}

The DWS angles from the reference and frequency interferometers report interesting information about how the optical bench is being distorted. The vertical angle on the reference interferometer --see Figure~\ref{RF_DWS_angles}-- is by far the most affected, followed by the one order of magnitude smaller vertical component of the frequency interferometer. 
Such an effect points to a mechanical deformation of the bench related with relative displacements on the Z axis. In addition, the pattern of signs $[-\,+\,+\,-\,+\,+\,]$ is coherent with the relative displacement readouts. Such scheme of signs is consistent with an eventual optical bench torsion along the Y axis, induced by stresses on Z, caused by the Z-component of the strut-elongation. Such torsion mode must expand/contract the optical path of the reference measurement more than the path of the other measurements and should bend the optical bench so its beams are tilt as observed. Under these circumstances, fluctuations affecting mainly the reference path on the optical bench would not be cancelled and therefore would be seen as a quite common distortion at all the other channels, as observed in~\ref{heaters_run}. Figure~\ref{ob_torsion} shows the torsion mechanism proposed.

The struts attaching the LCA to the spacecraft structure can be approximated to a $10\,{\rm cm}$-length part of Carbon-Fiber-Reinforced Plastic (CFRP) with $2.9\,{\rm cm}$ titanium end fittings at the edges~\footnote{These values correspond to a -Z strut at the OSTT campaign, while +Z struts were a factor 1.1 longer.}. Assuming average temperature increments for the CFRP part and the titanium end fittings of $3.5\,{\rm K}$ and $1\,{\rm K}$ respectively, and thermal expansion coefficients of $8.3\times 10^{-7}\,{\rm K^{-1}}$ and $8.6\times 10^{-6}\,{\rm K^{-1}}$, the projected free elongation on the Z axis represents a vertical displacement of $0.7\,{\rm \mu m}$. Since the LCA is a hyperstatic structure it is reasonable to expect a real displacement of at least one order of magnitude smaller, i.e $0.07\,{\rm \mu m}$, just a few times bigger than the measured perturbation at the displacement channels of $\approx 0.01\,{\rm \mu m}$. 

Another first order approximation can be found by comparing the observations against a pure bending mechanism of the optical bench around the X axis~\cite{bending_TN}. In this sense, considering a beam path of $D\approx 30\;{\rm cm}$ for the reference interferometer beams, the expected optical pathlength variation in front of a given beam angle $\theta$ at a photodiode can be expressed by
\begin{equation}
\Delta x =  \theta^2 \frac{D}{2}
\end{equation}
which turns to be $\Delta x \approx 0.6\;{\rm \mu m}$ for the DWS angle of $\theta \approx 2\;{\rm mrad}$ measured by the reference interferometer. This value is much larger (about a factor of 60) than the observed longitudinal measurements in ${\rm x_{12}}$ and ${\rm x_1}$. However we would expect this simple model to give a significant overestimate of the coupling through two independent effects. The first is that the actual distortion is significantly more complicated than the simple model, and the second is that the wavefront curvature of the optical beams will also significantly reduce the coupling factor.






\begin{figure}[t]
\begin{center}
\includegraphics[width=5.35cm,angle = -90]{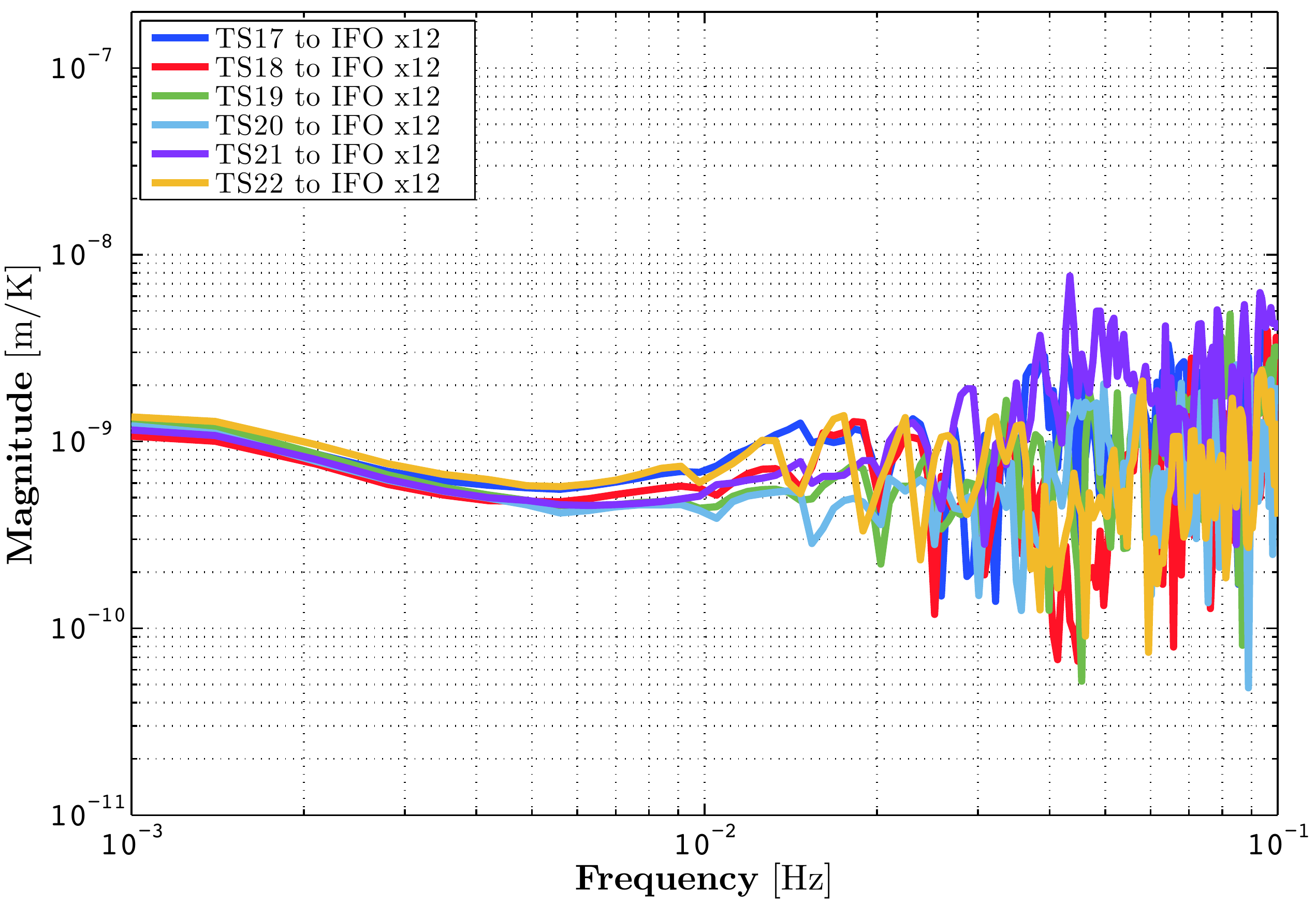}
\includegraphics[width=5.35cm,angle = -90]{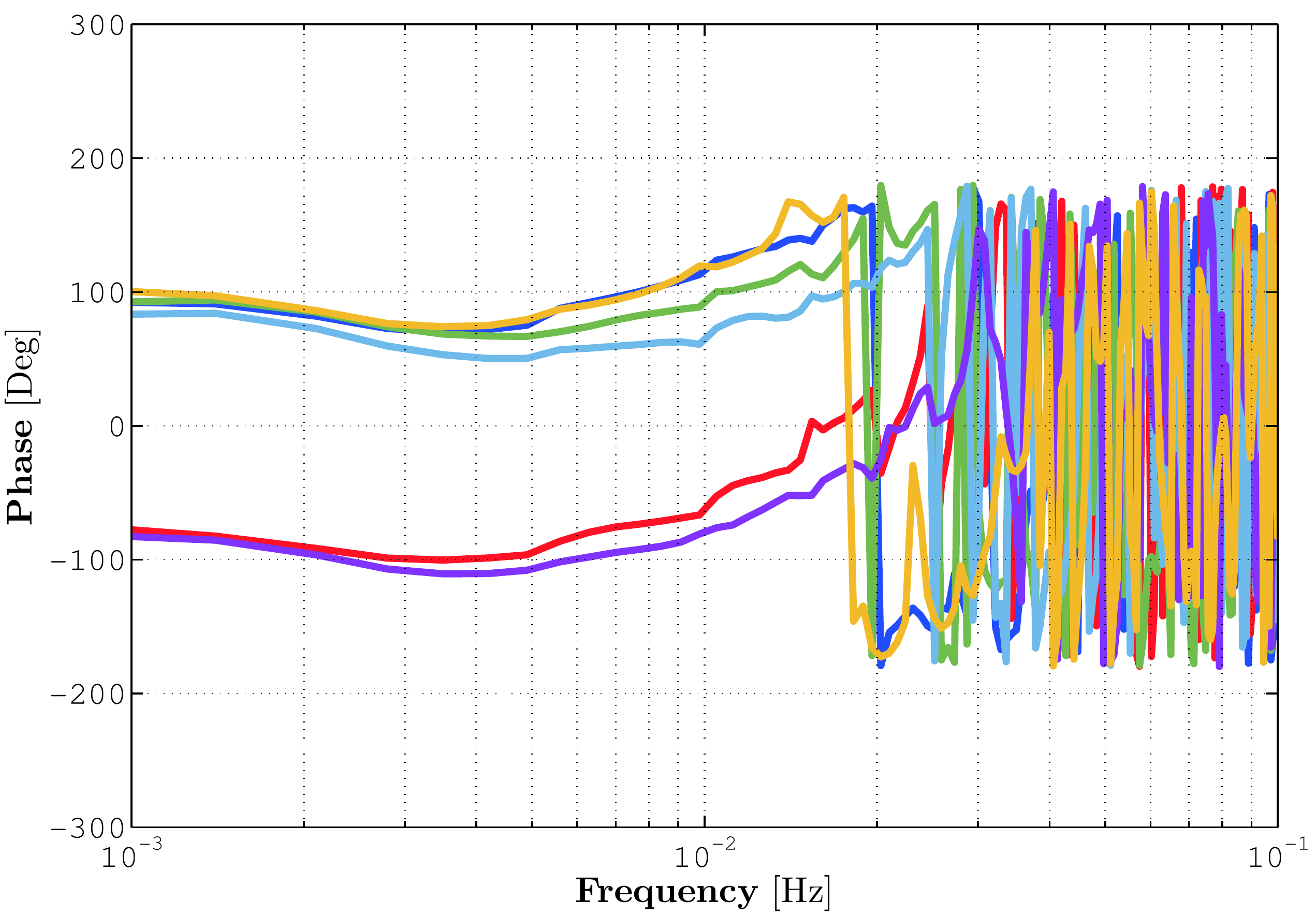}
\caption{Transfer functions from the temperatures at the Suspension Struts to displacement as measured by the interferometer measuring the distance between both test masses (${\rm x_{12}}$), magnitude ({\it left}) and phase ({\it right}).}
\label{strutsTS_TP15_3b_TFX1X12}
\end{center}
\end{figure}

\section{Projection of thermo-elastic induced phase noise}
\label{dist_charac}
Once the origin of interferometer response has been traced, we can proceed to quantify its contribution to the noise budget. We start by computing the individual transfer functions from the the temperature signals to the interferometer readouts, given by~\cite{Bendat}
\begin{equation}
H_{T_iX}(\omega) = \frac{S_{T_iX}(\omega)}{S_{T_iT_i}(\omega)}
\label{eq.transfer}
\end{equation}
where $T_i$ and $X$ are the Suspension Strut temperature sensor and the interferometer signals respectively, and $S_{T_iX}(\omega)$ and $S_{T_iT_i}(\omega)$ are the cross-power spectral 
density and the power spectral density. The transfer functions for each strut  is shown in Figure~\ref{strutsTS_TP15_3b_TFX1X12}.  The thermo-elastic coupling of the struts to the interferometer is of $10^{-9}\;{\rm m/K}$  throughout most of the LTP band. No significant differences between transfer functions are found since the geometry of the LCA is symmetric regarding all the suspension struts and the interferometer. The noise at frequencies higher than $70\,{\rm mHz}$ corresponds to the band where the ambient noise is lower than the electronic noise and the signals become uncorrelated. 

\begin{figure}[t]
\begin{center}
\includegraphics[width=6cm,angle = -90]{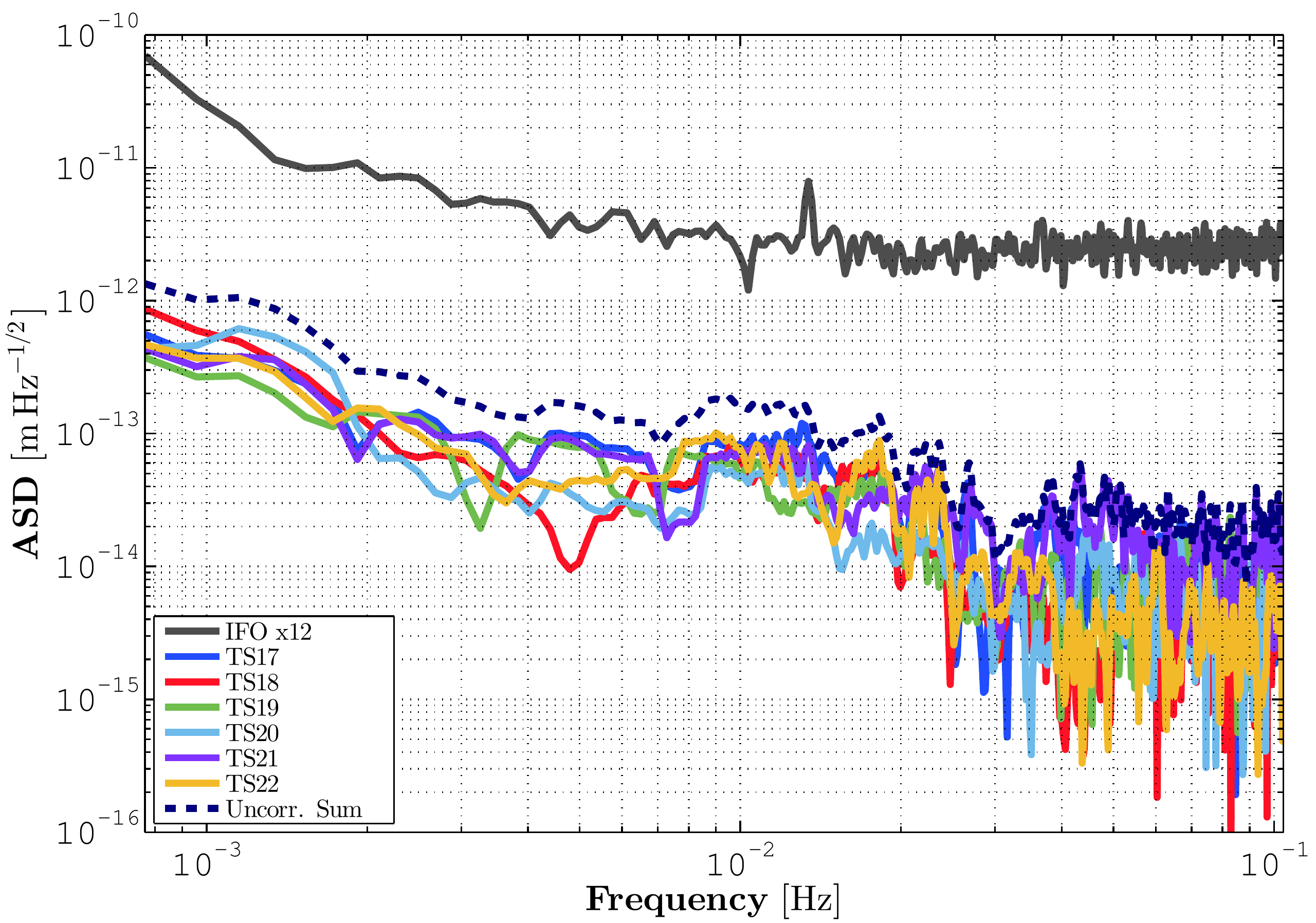}
\caption{Noise projection of thermo-elastic induced phase noise on the interferometer ${\rm x_{12}}$. The contribution from each suspension strut is shown together with the uncorrelated sum of all.}
\label{noiseProj1}
\end{center}
\end{figure}


The system linearity in front of this distortion is checked by comparing such transfer functions with respect to data from {\it Phase 3}. In this phase pulses of $500\,{\rm s}$ are applied simultaneously to TS17 and TS19 alternating with pulses to TS20 and TS22, producing a $2\,{\rm mHz}$ perturbation signal to the interferometer. The global transfer function recovered in this experiment matches with the coherent sum of the individual transfer functions for all the temperature sensors involved, meaning that, as expected, the thermoelastic strain mechanism keeps within its linearity.

\begin{figure}[t]
\begin{center}
\includegraphics[width=10cm,angle = -90]{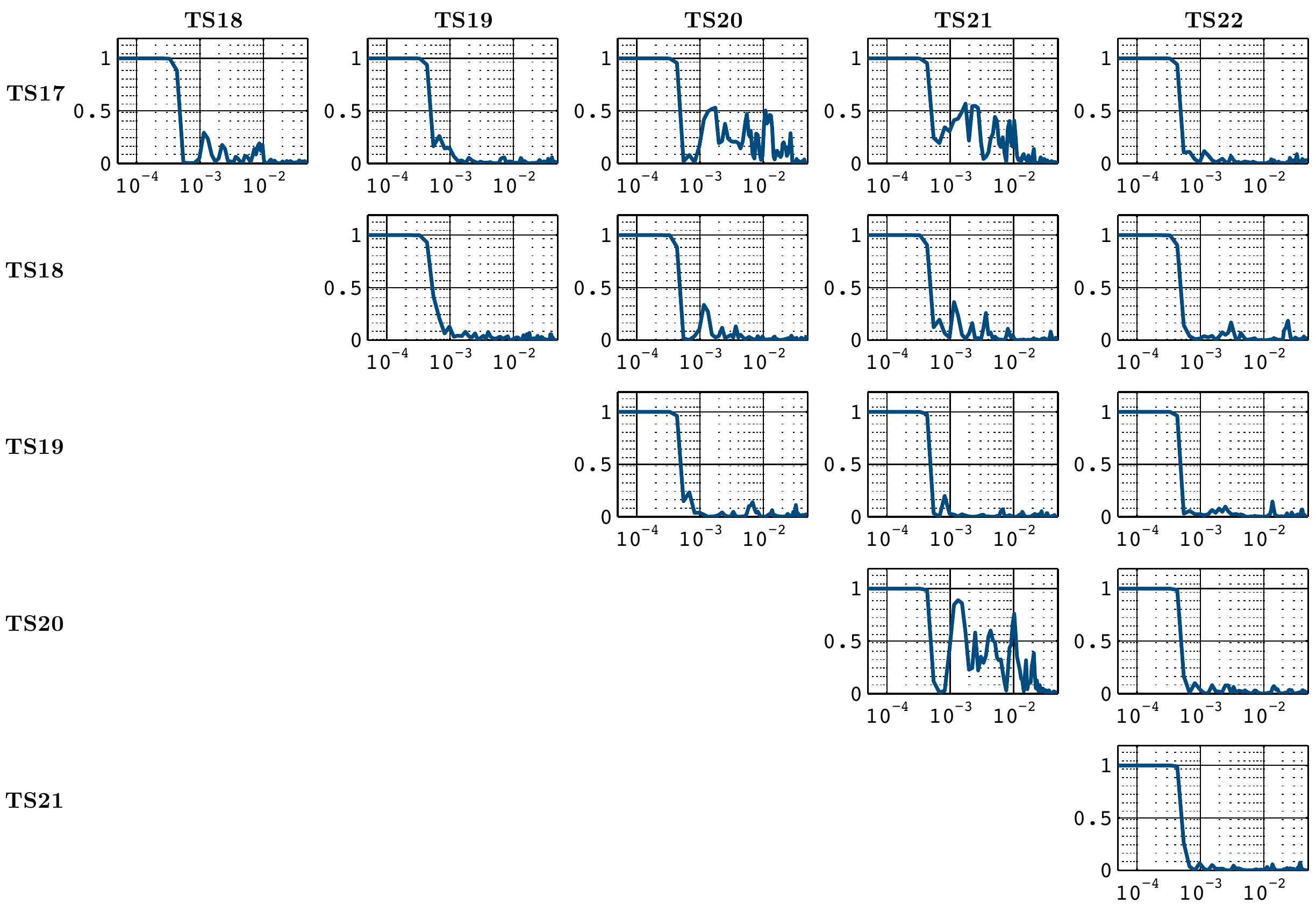}
\caption{Coherences between the different temperature sensors on the struts in the same timespan used for the computation of the temperature noise projection.}
\label{struts_coh}
\end{center}
\end{figure}

Following the previous notation, the thermo-elastic phase noise contribution to the total interferometer noise --known also as {\it noise projection}-- can be obtained by
\begin{equation}
S_{XX, \;T_i}(\omega) = H_{T_iX}(\omega) S_{T_iT_i}(\omega)
\label{eq.noiseproj}
\end{equation}
where to evaluate $S_{T_iT_i}(\omega)$ we now use a segment without heat injections, as the ones shown in Figure~\ref{ifoTP4}. The results ---in Figure~\ref{noiseProj1}--- show how the overall thermo-elastic phase noise contribution accounts approximately for $10^{-12}\;{\rm m/\sqrt{Hz}}$ at 1\,mHz, assuming that all contributions are added in an uncorrelated sum. Such assumption is feasible in the bandwidth of interest after discarding correlation through a coherence analysis between strut temperature sensors ---see Figure~\ref{struts_coh}. Although strong correlation dominates the band below $0.5\;{\rm mHz}$ and is present at some pairs of sensors (e.g. TS20 and TS21), the signals above $1\;{\rm mHz}$ are fairly enough uncorrelated. Still, Figure~\ref{noiseProj1} also shows the worse-case contribution of a hypothetical completely-correlated temperature noise distribution, where the temperature noise pattern considered is built by selecting the highest level of noise at each frequency for all the sensors involved. On the other hand, no differences were appreciated between contributions to the different interferometer readouts and also between the hot case and the cold case.

It is worth stressing that the obtained contribution is not representative of the flight operations since temperature stability is expected to be better in the Lagrange point L1 compared to the current experiment. However, the analysis reported here will be valid for the mission analysis. 

\section{Conclusions}
\label{conclusions}
We have presented first results for the thermal diagnostics sensors on the LISA Pathfinder spacecraft in space-like conditions. 
During the campaign, the temperature measuring system achieved an in-band sensitivity of $10^{-4}\,{\rm K}\, {\rm Hz}^{-1/2}$ in those parts not exposed to temperature drifts higher than $10^{-5}\,{\rm K/s}$. The electronic noise of the measurement system kept the noise floor down to $10^{-5}\,{\rm K}\, {\rm Hz}^{-1/2}$. 
Among other thermal experiments, a set of heat injections were applied at each Suspension Strut anchoring the LTP Core Assembly at the satellite's mechanical support structure for the LTP, which is acting as well as thermal shield. This temperature modulation allowed us to experimentally determine the coupling of temperature fluctuations on the struts to the interferometer onboard for the first time. We have estimated a nearly flat transfer function throughout the measuring bandwidth with magnitude $10^{-9}\;{\rm m/K}$. 

At the same time, we have allocated the contribution arising from temperature fluctuations in the struts to be $10^{-12}\;{\rm m/\sqrt{Hz}}$ at 1\,mHz, around a factor 30 below the main interferometer measurement floor noise, for the campaign conditions (which in turn was a factor of approximately 3 better than required for flight). Although the obtained results are related to the particular temperature environment of the campaign, the methodology described here is readily applicable to the in-flight case.

We have also investigated the origin of the coupling between temperature fluctuations and interferometer response to conclude that the effect has a thermo-elastic origin, discarding direct temperature gradients on the optical bench as the cause of the observed phase response. A thermo-elastic mechanism consisting of an optical bench torsion around the Y axis, induced mainly by the vertical component of the stress caused by the elongation of the heated strut is consistent with the observations.

Coupling through other thermal sensitive locations, as for instance the electrode housing surrounding the test mass can be disregarded since they were not present in the current setup. The temperature-induced pathlength variation at the Optical Window~\cite{ow_paper} has not been assessed since, though the Optical Windows were present in the setup in a thermally almost representative mount with respect to the optical bench, no mission-planned experiments were performed on them. These effects, which will be the aim of the thermal experiment during flight operations, are being end-to-end simulated~\cite{ssm_exp} and tested on-ground through other setups~\cite{pendulum_trento, ow_paper}.



\appendix
%

\section{Identification and fit of the ADC non-linear error}
\label{adcerror}
The bumps and sudden falls that appear in the power spectra of some temperature sensors, in the LPF band --see Figures~\ref{ifoTP4}--, are individually associated to integral non-linearities of specific bits of the 16-bit Analog-to-Digital Converter (ADC). Their affected frequencies depend on the signal drift. Each bit contribution --as a voltage error-- is frequency-dependent and can be expressed as~\cite{wadgy_adc}:


\begin{equation}
| q_k (\omega) | = \Delta \sum^{\infty}_{n=-\infty} \frac{\sin \frac{n \pi \epsilon_k}{2^{k+1}\Delta}}{n} \frac{\sin \frac{n \pi}{2}}{\sin \frac{n \pi}{2^{k+1}}} \delta \left(\omega - \frac{\pi n |b|}{2^k \Delta} \right)
\label{eq.adcerror}
\end{equation}

where:
\begin{itemize}
\item $b$ is the voltage slope, in ${\rm V/s}$.
\item $\Delta$ is the ADC quantization step in ${\rm V}$. 
\item $k$ is the binary digit, increasing from 0 to 15, i.e. from the least significant bit to the most significant.
\item $\epsilon_k$ is the relative error associated to each bit, as a fraction of $\Delta$.
\end{itemize}

\begin{figure}[t!]
\begin{center}
\hspace{-3cm}
\begin{minipage}[h]{0.3\textwidth}
\includegraphics[width=5.5cm,angle = -90]{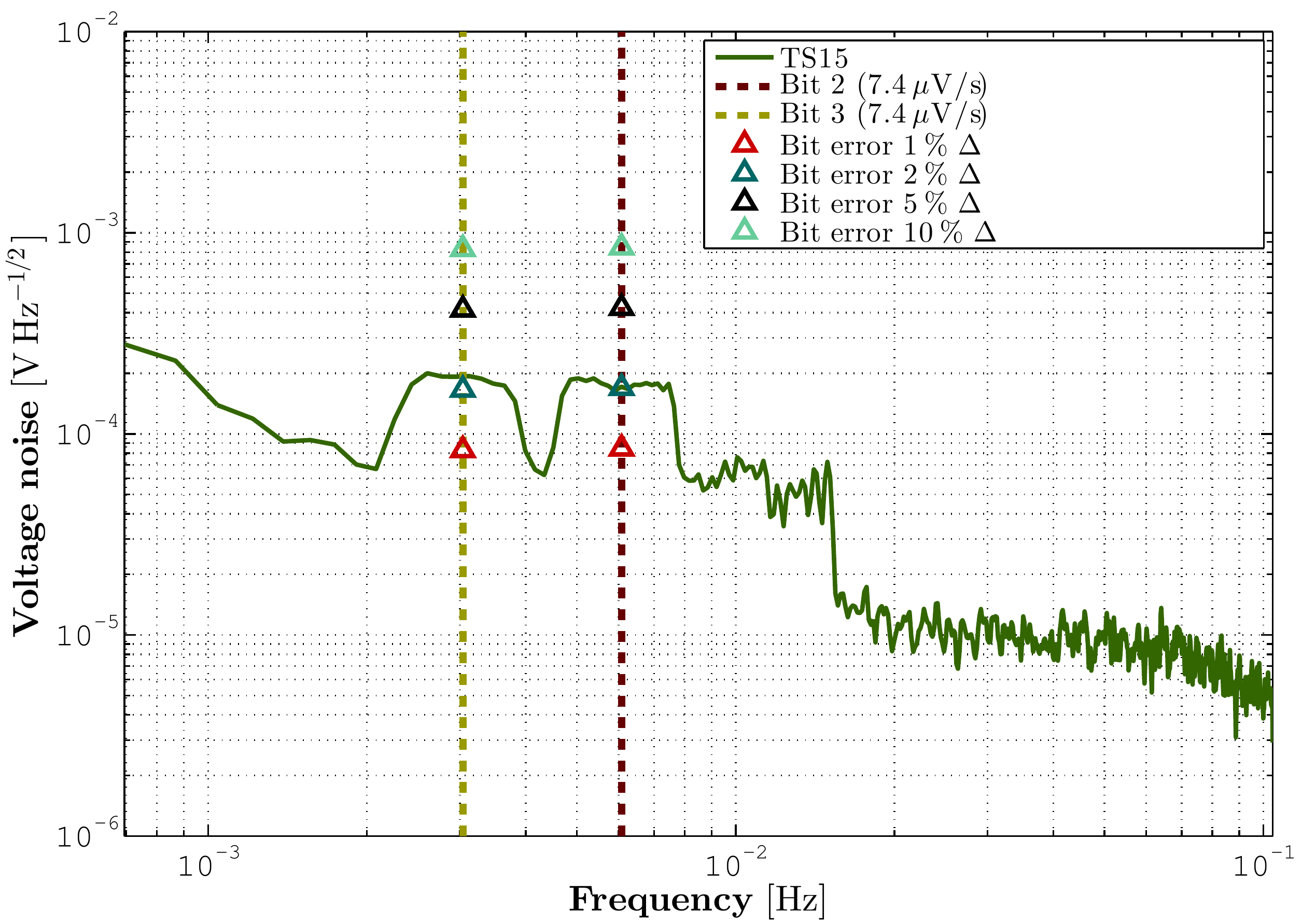}
\end{minipage} \hspace{0.18\textwidth}
\begin{minipage}[h]{0.3\textwidth}
\vspace{-0.52cm}
\includegraphics[width=5.85cm,angle = -90]{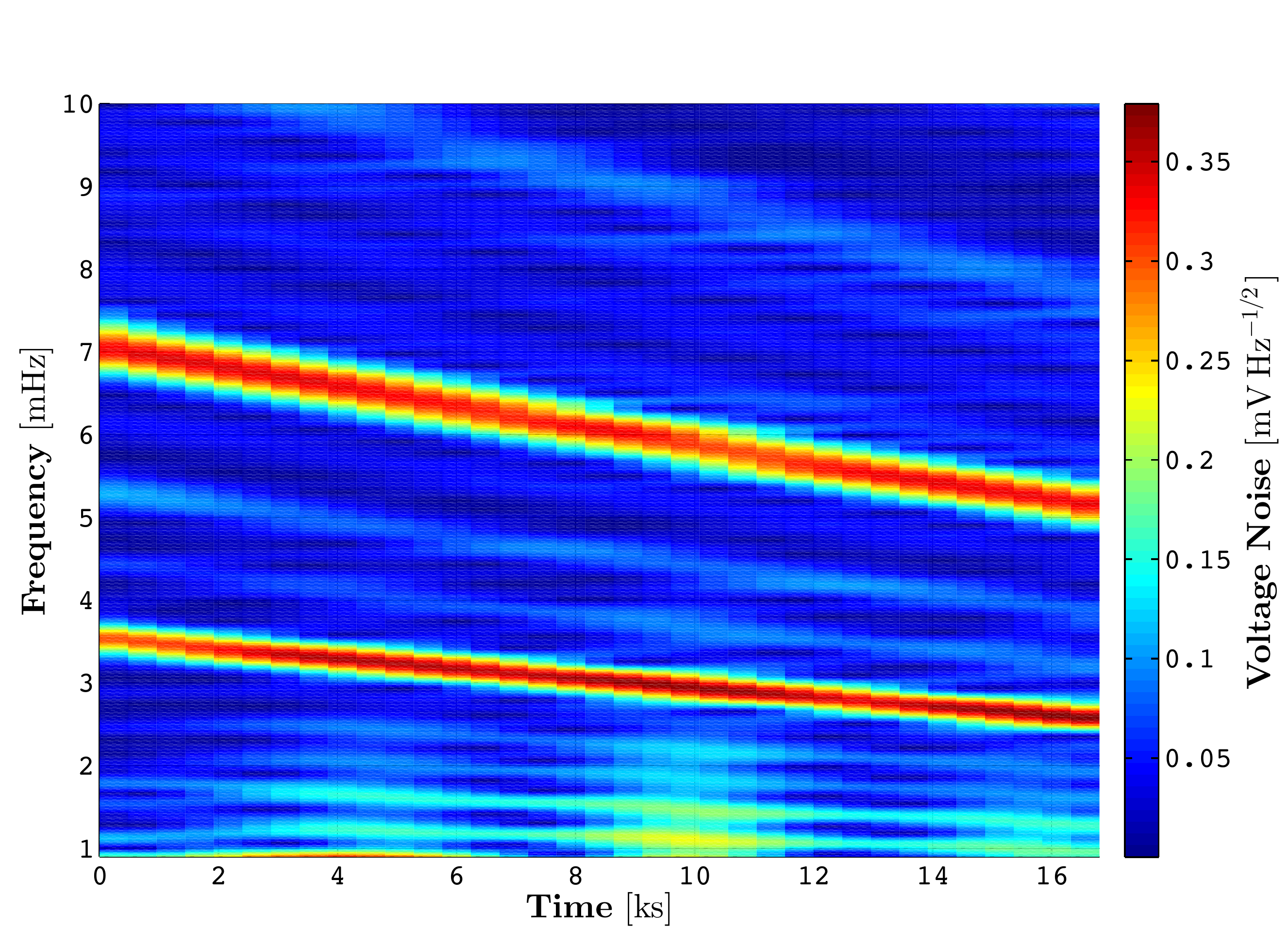}
\end{minipage}
\caption{{\it Left}: PSD of the voltage noise at TS15 at specific segment where the ADC non-linear effects create two bumps around $3\, {\rm mHz}$ and $6\, {\rm mHz}$ caused by bit errors at bits 3 and 2 respectively in front of a voltage drift varying around $7.4\, {\rm \mu V/s}$. The bit errors provided in percentage of $\Delta$ correspond to the theoretical values computed with Equation~\ref{eq.adcerror} for bits 2 and 3 with the measured average drift --only the error amplitudes for the fundamental frequencies are shown. The exact amounts of $\Delta$ mismatch for each bit cannot be determined since depend as well on the environment noise at the affected bands, which are shaped by the error itself. {\it Right}: Spectrogram of the voltage during chosen segment. The reddish lines show the affected frequencies varying within the time, caused by the variation of temperature drift.}
\label{adcnoise}
\end{center}
\end{figure}

Nevertheless, a drifting {\it noisy} input signal instead of a {\it pure} voltage drift --which is the case of the studied temperature measurements-- strongly dumps the high frequency contributions at~\ref{eq.adcerror}. Such an effect helps to reduce the impact of the nonlinear effects at high frequencies with the tradeoff of spreading their energy across the whole spectrum. Inducing out-of-band dither noise
to the signal prior to the ADC input is often used to increase ADC sensitivity at specific bands~\cite{pep_adc}.

In the case at Figure~\ref{adcnoise} the affected  frequencies clearly oscillate close to $3\, {\rm mHz}$ and $6\, {\rm mHz}$, which is consistent with a bit mismatch at bits 2 and 3 of the ADC in front of a mean drift of $6\, {\rm \mu K/s}$, considering a system sensitivity of $1.35\, {\rm V/K}$. The corresponding temperature noise level introduced, around $ 1.5 \times 10^{-4}\,{\rm K}\, {\rm Hz}^{-1/2}$, clearly exceeds the ambient level by a factor two. Non-linear effects at the highest side of the spectrum are self-mitigated by the signal noise.



Limits to the temperature drift can be obtained from this analysis. Since the bumps noise amplitude of this error depend only on the $\Delta$ parameter which cannot be modified, the highest-frequency bump must be shifted down, out of the bandwidth of interest. In this sense, in order to reduce its highest frequency affected ($8\, {\rm mHz}$) to $1\, {\rm mHz}$, the temperature drift must be therefore limited to 1/8 of the measured, around $0.75\, {\rm \mu K/s}$.



\section*{Acknowledgements}
The authors acknowledge support from Project AYA2010-15709 of {\it Plan Nacional del Espacio} of the Spanish Ministry of Science and Innovation (MICINN) and support by  Deutsches Zentrum f\"{u}r Luft- und Raumfahrt (DLR) with funding of the Bundesministerium f\"{u}r Wirtschaft und Technologie project reference No 50OQ0501, and support from the UK Space Agency, and the University of Glasgow. Airbus Defence and Space GmbH, Friedrichshafen, Germany, planned and conducted the LTP thermal experiments based on ICE-CSIC/IEEC plans and provided the test data obtained at occasion of the LISA Pathfinder Spacecraft On-Station Thermal Test. This thermal test was conducted by Airbus Defence and Space UK as the spacecraft responsible. ESA supported by several national space agencies is responsible for the entire LISA Pathfinder program. MN acknowledges a JAE-DOC contract from CSIC (co-funded by the ESF) and support from the EU Marie Curie CIG 322288.

\end{document}